\documentclass[aps,pre,reprint]{revtex4-1}

\usepackage{amstext}
\usepackage{amsmath}
\usepackage{amssymb}
\usepackage{amsfonts}
\usepackage{units}

\usepackage{graphicx}
\usepackage{color}
\usepackage[normalem]{ulem} 

\usepackage{microtype}
\usepackage[pdfborder={0 0 0}]{hyperref} 
\usepackage{mathtools}

\renewcommand{\vec}[1]{\boldsymbol{#1}}

\definecolor{Red}{rgb}{0.9,0.0,0.1}
\definecolor{Blue}{rgb}{0.1,0.0,0.9}
\definecolor{Green}{rgb}{0.0,0.5,0.0}
\definecolor{Darkblue}{rgb}{0.22,0.33,0.64}

\begin{document}

\author{Christian Wischnewski} 
\affiliation{Department of Physics, Technische Universit\"{a}t Dortmund, 44221 Dortmund, Germany}

\author{Elena Zwar} 
\author{Heinz Rehage} 
\affiliation{Department of
 Chemistry, Technische Universit\"{a}t Dortmund, 44221 Dortmund, Germany}

\author{Jan Kierfeld} 
\email{jan.kierfeld@tu-dortmund.de}
\affiliation{Department of Physics, Technische Universit\"{a}t Dortmund, 44221
 Dortmund, Germany}

\title{{Strong deformation of ferrofluid-filled elastic
 alginate capsules in inhomogenous magnetic fields}}

\begin{abstract}
 We present a new system based on alginate gels for the encapsulation of a
 ferrofluid drop, which allows us to create millimeter-sized elastic capsules
 that are highly deformable by inhomogeneous magnetic fields. We use a
 combination of experimental and theoretical work in order to characterize
 and quantify the deformation behavior of these ferrofluid-filled capsules.
 We introduce a novel method for the direct encapsulation of unpolar
 liquids by sodium alginate. By adding 1-hexanol 
 to the unpolar liquid, we can
 dissolve sufficient amounts of $\mathrm{CaCl_2}$ in the resulting 
  mixture for ionotropic gelation of sodium alginate.
The addition of polar alcohol molecules 
allows us to encapsulate a ferrofluid as a single phase 
  rather than an emulsion without impairing ferrofluid stability. 
  This encapsulation method 
increases the
 amount of encapsulated magnetic nanoparticles resulting in high deformations
 of approximately 30\% (in height-to-width ratio) in inhomogeneous magnetic
 field with magnetic field variations of $50\,\mathrm{mT}$ over the size of
 the capsule. This offers possible applications of capsules as actuators,
 switches, or valves in confined spaces like microfluidic devices. We
 determine both elastic moduli of the capsule shell, Young's modulus and
 Poisson's ratio, by employing two independent mechanical methods, spinning
 capsule measurements and capsule compression between parallel plates. We
 then show that the observed magnetic deformation can be fully understood
 from magnetic forces exerted by the ferrofluid on the capsule shell if the
 magnetic field distribution and magnetization properties of the ferrofluid
 are known. We perform a detailed analysis of the magnetic deformation by
 employing a theoretical model based on nonlinear elasticity theory. Using
 an iterative solution scheme that couples a finite element / boundary
 element method for the magnetic field calculation to the solution of the
 elastic shape equations, we achieve quantitative agreement between theory
 and experiment for deformed capsule shapes using the Young modulus from
 mechanical characterization and the surface Poisson ratio as a fit
 parameter. This detailed analysis confirms the results from mechanical
 characterization that the surface Poisson ratio of the alginate shell is
 close to unity, that is, deformations of the alginate shell are almost area
 conserving.
\end{abstract}

\maketitle

\section{Introduction}

Ferrofluids contain dispersed magnetizable nanoparticles, which 
are long-time stable and exhibit superparamagnetic behavior
\cite{Rosensweig1985}. Ferrofluids can be actuated by magnetic 
fields and have various 
technical and medical applications \cite{Voltairas01,Holligan03,Liu07}. 
In many applications it is of interest to prevent a ferrofluid from
interaction with its environment, especially considering its corrosive effects
on metals. This can be
achieved by encapsulation of a ferrofluid drop with a thin protective 
 shell made, for example, from a gel or soft elastic material. The
result is an elastic capsule with a stable ferrofluid droplet inside. 
The elastic
shell protects the inner fluid from direct interaction with the environment
but it can be deformed by various external forces 
\cite{Fery04,fery_mechanical_2007}. 
Elastic capsules are stable under uniform pressure up to a buckling
threshold \cite{Zoldesi08,Gao01,Knoche11}.
Buckling indentations can also be induced by 
point forces \cite{Fery04,Vella2012} and capsules can be deformed
in hydrodynamic flow
\cite{Barthes-Biesel2011}. 
Other deformation techniques include compression between parallel plates 
\cite{carin_compression_2003,fery_mechanical_2007} or in a spinning 
drop apparatus \cite{pieper_deformation_1998}, or pendant capsule 
elastometry \cite{Knoche13,Hegemann2018}. 
These mechanical deformation techniques can be employed to determine
the elastic moduli of capsule shells. 
Whereas these deformation techniques require direct mechanical interaction
with the capsules, 
deformation can also be driven by external electric or
magnetic fields if the capsules are filled 
by a dielectric liquid or ferrofluid 
\cite{Degen08,Karyappa2014,Wischnewski2018} 
or if the shell itself contains magnetizable particles
\cite{Loukaides2014,Seffen2016}.
In particular, we demonstrate in this paper that 
magnetic actuation and mechanical characterization of a capsule 
filled with a ferrofluid is possible 
by deformation in a magnetic field.

The aim of this work is the encapsulation of a stable 
ferrofluid drop with high magnetic nanoparticle concentration
in order to achieve strong deformation 
 in an external magnetic field and a quantitative 
understanding of the deformation behavior by comparison to
theoretical predictions based on nonlinear elasticity theory. 
Encapsulation of stable
 ferrofluids was achieved in Refs.\
 \citenum{neveuprin1993,shen_properties_2003}, for
 example, for magnetic resonance 
imaging \cite{shen_properties_2003}. Ferrofluid 
concentration inside the capsules were low, however, such that there are
 no reports on actuation and deformation properties in magnetic 
fields. On the other hand, magnetic deformation of capsules filled
with magnetic particles was reported in 
Refs.\ \citenum{degen_stimulated_2008,degen_magneto-responsive_2015}; these 
magnetic particles, however,  aggregated and 
 did not remain in a stable ferrofluid state.
 The aim of this
 work is a reliable capsule system, into which 
 a stable ferrofluid can be loaded at high concentration such that it
 can strongly deform in magnetic fields, which is important 
for possible applications of capsules as actuators, 
switches, or valves.

We present a new method 
to achieve stable ferrofluid encapsulation at high concentrations.
 Generally, encapsulation of magnetic nanoparticles in a stable 
dispersion in the liquid core is a difficult task.
Here we use alginate gel as shell material. 
Because of its mild gelation conditions, non-toxicity and
biocompatibility, it is a widely used encapsulation material
\cite{martins_oil_2015,degen_magneto-responsive_2015,martins_oil_2017,Lee2012,Leong2016,Prakash2007}
Sodium alginate is soluble in water, and ionotropic gelation
takes place upon contact with bi- or trivalent cations
\cite{ouwerx_physico-chemical_1998,morch_effect_2006};
 usually, calcium is used for the gelation. 
Magnetic nanoparticles, which are sterically stabilized by amphiphiles such 
as lauric acid or electrostatically stabilized, destabilize 
and flocculate in the presence of calcium ions 
 if both ions and nanoparticles 
are in the aqueous phase in the liquid core of the 
capsule \cite{degen_magneto-responsive_2015}.
To reduce this effect oil-soluble
magnetic particles can be used, 
which are encapsulated in an oil in water emulsion,
where calcium is solved in the aqueous emulsion phase
 \cite{martins_oil_2017}. In previous work,
 we showed that encapsulation of a stable
ferrofluid is possible with this technique and that a significant deformation
in magnetic fields can be achieved \cite{zwar_production_2018}.
The deformation 
was, however, limited to approximately 10\% because 
magnetic nanoparticles are only contained in the oil emulsion phase, 
whereas a certain amount of unmagnetizable aqueous emulsion phase is needed to 
solve calcium ions.
In order to further increase the magnetic nanoparticle content in the liquid
core of the capsule we use 1-hexanol as a solvent
with high polarity but low water-solubility to dissolve
significant amounts of calcium chloride. 
We mix 1-hexanol with chloroform as a second solvent because the
magnetite nanoparticles are not dispersable in pure 1-hexanol. 
Chloroform also increases the density of the liquid core 
which improves experimental handling of the capsules.
The
 general principle of using 1-hexanol as an additive can be extended to a
 variety of oils. The advantage of direct oil encapsulation concerning the
 encapsulation of ferrofluids is the higher magnetic content, which leads to
 higher deformation in the same magnetic field. 
 In emulsion-based systems the need for a
 second phase reduces the overall nanoparticle concentration and thus the
 reaction to magnetic fields.

The ferrofluid-filled capsule represents a magnetic dipole. In order to
achieve strong deformation we use inhomogeneous magnetic fields, which are
easily realizable and result in a net force onto the ferrofluid-filled
capsule. We use this net force to deform the capsule by pushing the
 flexible particle against the bottom wall of the cuvette.
In principle,
deformation and shape transitions are also possible in homogeneous magnetic
fields, which tend to stretch the capsule in order to increase the dipole
moment \cite{Wischnewski2018}. For a quantitative understanding of the
deformation behavior we measure the magnetic field distribution, calculate the
magnetic properties of the ferrofluid from the nanoparticle size distribution
and compare the experimental capsule shapes to theoretical shapes calculated
using nonlinear elasticity theory by a coupled finite element method (FEM) 
for the
elastic problem and a boundary element method (BEM) for the magnetic field
calculation. This enables us to also obtain additional information not easily
accessible in experiments such as a detailed picture of the magnetic field
distribution and the elastic stress distribution in the capsule, which are
important, for example, to predict capsule rupture because of magnetic
deformation.

\section{Materials and methods}

\subsection{Preparation of the magnetic nanoparticles}
The magnetite nanoparticles ($\mathrm{Fe_3O_4}$) used in this work were
synthesized following a procedure by Sun \textit{et al.}
\cite{sun_monodisperse_2004}. These particles are crystalline, and usually
show a narrow size distribution around an average of 6\,nm diameter. The
small size is important for the stability of the resulting ferrofluid.
Additional stabilization is provided by the surfactants oleic 
acid and oleyl amine which hinder agglomeration.

\subsection{Preparation of capsules}

We encapsulate a mixture of chloroform and 1-hexanol (7:3) containing
dispersed magnetite nanoparticles with a mass concentration 
$c_m = 516\,\mathrm{g}/\mathrm{mol}$ in an alginate gel shell. 
 Initially, calcium chloride
is dissolved in the 1-hexanol/chloroform 
mixture (by volume 3:7)
 in order to perform the alginate gelation.
Two additional surfactants stabilizing the nanoparticles,
oleic acid and oleylamine, also accelerate the gelation 
and are added with 1\%$_V$ (volume percent) each to improve the process.

\begin{figure}[htb]
 \centering
\includegraphics[width=.35\textwidth]{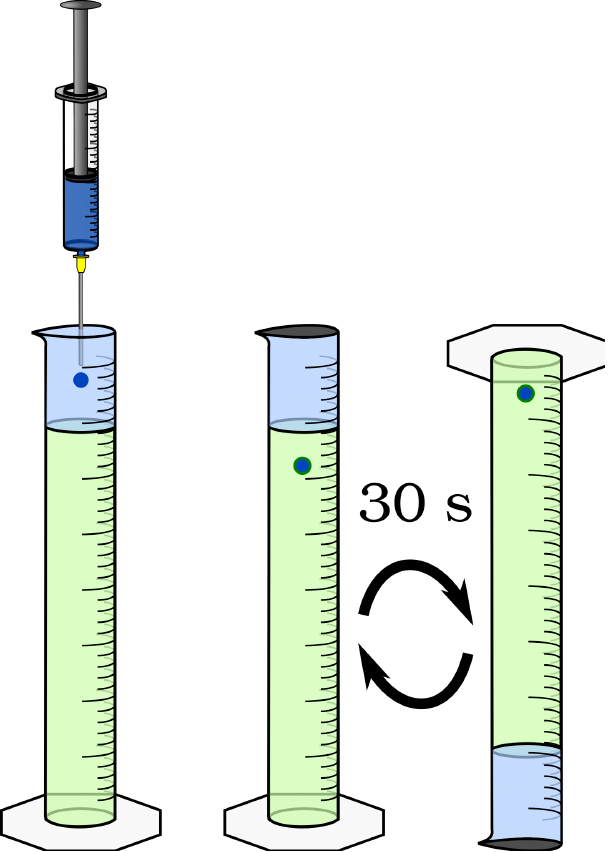}
\caption{Procedure for the production of ferrofluid-filled capsules,
  dark blue = 1-hexanol/chloroform mixture containing calcium chloride and 
 magnetic nanoparticles, light blue = water, 
  green = sodium alginate solution}
\label{fig:KapsHerstell}
\end{figure}

The high chloroform content requires to adjust 
the capsule preparation process, which is normally done 
 by dripping one component,
either sodium alginate or calcium chloride solution, directly (from air) 
into the other
liquid. Because of the high chloroform content of 70\%$_V$, 
dripping of the 1-hexanol/chloroform mixture 
containing the calcium chloride
from air into the alginate solution is not feasible because the mixture
spreads on the interface. 
Instead we first overlay a cylinder which is filled to 7/8 
with sodium alginate solution (1\% by weight) 
with distilled water and form a droplet of the 1-hexanol/chloroform mixture 
in the water layer using a capillary, see Figure \ref{fig:KapsHerstell}). 
The droplet then falls through the interface between water and alginate 
solution without spreading. 
Within the alginate solution, the calcium chloride dissolved 
in the 1-hexanol/chloroform droplet starts the gelation.
In order to avoid contact with the glass surface a seal 
(shown in grey in Figure \ref{fig:KapsHerstell})
can be placed over the opening and the cylinder can be turned for 30\,s.

The capsules are washed with water to stop the polymerisation and placed
in saturated sodium alginate solution. This lowers the elastic moduli
of alginate systems and thus lead to capsules that are easier to deform
\cite{zwar_production_2018}.

\subsection{Experimental setup for mechanical characterization}

For mechanical characterization of the capsule shells we use 
two methods, compression between parallel plates 
and deformation in a spinning drop apparatus.
Combination of the results from both measurements 
 will enable us to determine 
both the two-dimensional Young
modulus $Y_{2\text{D}}$ and the surface Poisson ratio $\nu$.

In the compression method 
a capsule is placed between two
parallel plates and the force is measured as a function of the
displacement. 
A sketch of the method is shown in the Supporting Information. 
Compression between parallel plates was performed
 with the DCAT11 tensiometer (DataPhysics Instruments GmbH) 
with the software SCAT. The compression speed
was set to $0.02\,\mathrm{mm/s}$.

In a spinning drop apparatus,
a capsule is
placed in a capillary filled with a liquid of higher density $\rho$ 
 and monitored with a camera (see Supporting Information). 
During rotation of the capillary, the capsule
moves to the horizontal axis of the capillary and deforms.
The deformation is measured as a function of angular rotation frequency.
For the spinning capsule experiments, we used
the SVT 20 of the DataPhysics Instruments
GmbH. We used Fluorinert 70 (FC 70) as outer
phase because of its high density. The initial
undeformed (quiescent) capsule state was recorded at
2000 rpm.

Capsule radii were determined by image analysis of capsule photos
(see Supporting Information). 
 For the shell thickness this image analysis
could not be used because of the small thickness. Thus, scanning electron
microscopy (SEM) measurements were performed to estimate the shell
thickness.
From the images shown in the Supporting Information, it can also 
be concluded that the nanoparticles were not incorporated in the shell.

\subsection{Experimental setup for magnetic deformation}
 
\begin{figure}[h]
 \centering
\includegraphics[width=.48\textwidth]{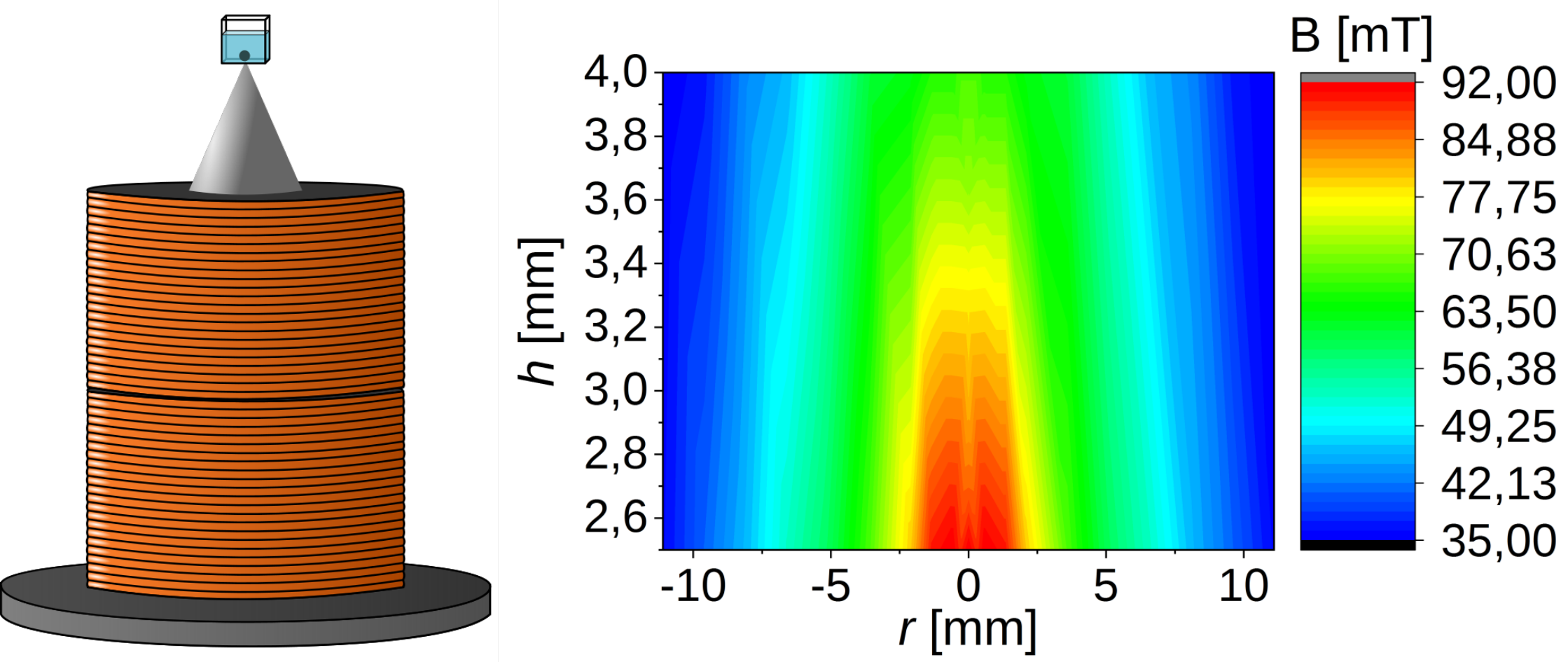}
 \caption{Experimental setup and measured flux density for the magnetic
  deformation of capsules with a current of $I=2\,$A. }
 \label{fig:Aufbau}
\end{figure}

The ferrofluid filled capsule is placed at the bottom of a nonmagnetic
cuvette. This cuvette was placed right above the conical iron core of a
coil. This way, the capsule was as close as possible to the tip of the cone
(the bottom of the capsule was about 2.6\,mm above the iron core) and exceeded
a maximum field strength. In Figure \ref{fig:Aufbau}, the experimental setup
for the deformation of capsules in magnetic fields is shown.

\subsection{Elastic and magnetic model for deformation}

In order to model the magnetic deformation of a capsule we need 
an elastic material model for the capsule shell and 
a magnetic model for the calculation of the magnetic forces. 
To calculate the shape of a deformed capsule, we use a small strain
nonlinear shell theory with a Hookean elastic law
\cite{Knoche11,Knoche13,Knoche14}. In our model, we assume 
a very thin, effectively two-dimensional elastic
shell and rotational symmetry. 
We also assume that the capsule attains a perfectly spherical 
stress-free reference shape 
with rest radius $R_0$ during the polymerization process.
All these assumptions hold to a good approximation in the experimental 
realization. 
 Once the elastic shell is polymerized, there is no exchange
of fluid between the inner phase and the outer phase possible anymore,
at least under the employed  experimental conditions and 
 during magnetic deformation.
So the volume $V_0$ of the capsule remains constant. 
The capsule is then deformed by a
hydrostatic pressure caused by the density difference $\Delta\rho$ between the
inner phase and the outer phase. 
In the presence of an additional magnetic field,
the magnetic stress caused by the ferrofluid deforms the capsule as well.
We consider capsules which sink to the bottom of a cuvette where 
also the magnetic deformation takes place, and both gravity and magnetic 
forces press the capsule against the bottom wall of the cuvette. 
In a steady state without motion, the fluid can only exert normal forces
 on the surface. The normal forces caused by the ferrofluid are given by
\cite{Rosensweig1985}
\begin{align}  \label{eq:stress_tensor}
 \begin{split}
	f_m(r,z) &= \mu_0 \int\limits_0^{H(r,z)}M(r,z)\text{d}H'(r,z) \\
	&\quad+ \frac{\mu_0}{2}M_n^2(r,z)
	\end{split}
\end{align}
($M$ and $M_n$ are absolute value and normal component of the 
magnetization ${\bf M}$, $H$ the magnetic field, and we use cylindrical 
coordinates $r$ and $z$; for a ferrofluid we can safely assume 
that ${\bf M}$ and ${\bf H}$ are parallel). 
The deformed shape of the capsule can be calculated by using a system of six
nonlinear differential equations, the shape equations, 
see Ref.\ \citenum{Knoche11} for
a derivation and the Supporting Information for more details.

There are two material parameters, the two-dimensional Young modulus 
 $Y_{2\text{D}}$ and the surface Poisson ratio $\nu$, that enter the
theoretical model. 
These parameters are obtained from the mechanical characterization of the 
capsule by plate compression and spinning capsule deformation
as explained below in the Experiments and Results section. 
Moreover,  the rest radius $R_0$ of the
initially spherical capsule and the thickness $h$ of the shell are needed. 
The thickness $h$ determines the bending modulus via 
$E_B = Y_{2\text{D}}h^2/12(1-\nu^2)$ and $R_0$ the initial rest shape 
with respect to which elastic strains and stresses are obtained 
and the fixed volume $V_0$. 
Moreover,  the density difference $\Delta \rho$ between
the interior and exterior liquid phases is needed 
for the hydrostatic pressure.

 In addition, the effective interface tension $\gamma$ between the inner
 liquid phase and the elastic shell and between the shell and the outer
 liquid phase has to be estimated, but is very difficult to be measured. 
 Generally we expect solid-liquid surface tensions to be smaller 
 than liquid-liquid surface tensions \cite{Mondal2015}.
 Because the shell can still contain pores leading to liquid-liquid 
 contact, we 
  expect the interface tension $\gamma$ to be lower than but similar to 
 the interfacial
 tension between outer and inner liquid.
   The interfacial tension of a similar system without elastic shell
 and without nanoparticles inside was determined to be
 $14.6\,\text{mN}/\text{m}$ by pendant drop tensiometry. The presence of
 surfactants from the synthesis of the nanoparticles
 as well as the elastic shell should lower the interface
 tension below this value. In addition, it should be mentioned that the
 interfacial tension of the ferrofluid is probably increasing for
 increasing external magnetic field as Afkhami {\it et al.} observed 
 \cite{Afkhami2010}.
  We 
 expect $\gamma \approx 10\pm 4\,\text{mN}/\text{m}$ to be a valid average
 value.

The parameter values that are used 
for the numerical calculation of capsule shapes
are summarized in Table \ref{tab:Parameters}.
 The surface Poisson ratio deviates slightly from the value 
 measured in the experiment (eq \ref{eq:Ynu}).
This will be discussed below in the Results section.

\begin{table}
\begin{tabular}{lcc}
Name & Symbol & Value \\
\hline
2D Young's modulus & $Y_{2\text{D}}$ & $0.186\,\text{N}/\text{m}$ \\
Surface Poisson ratio & $\nu$ & 0.946\\
Shell thickness & $h$ & $3\,\mu\text{m}$\\
Radius & $R_0$ & $0.903$ ... \\
 & & $1.044\,\text{mm}$\\
Density difference & $\Delta\rho$ & $0.24\,\text{g}/\text{cm}^3$ \\
Surface tension & $\gamma$ & $10\,\text{mN}/\text{m}$
 \end{tabular}
\caption{Numerical parameters}
\label{tab:Parameters}
\end{table}

Finally, the exact distribution of the applied 
inhomogeneous magnetic field and 
the magnetic properties of
the ferrofluid, that is, its magnetization curve $M=M(H)$ 
 have to be known as well in order to calculate the 
magnetic forces (eq \ref{eq:stress_tensor}), 
see following sections.

\subsection{Magnetic field}
\label{sec:field}

We used a Hall probe to measure the magnetic field depending on the position
and the applied current $I$ in the coil (see Figure \ref{fig:Aufbau}). 
 The capsules are very small in
relation to the size of the coil and the conical iron core, so the radial
component of the magnetic field can be neglected. In addition, the
$z$-component of the field is nearly constant in radial direction 
over the capsule size and thus only depends on the
$z$-coordinate. 
In the numerical calculations we use a fit to the 
measured external magnetic field with a Langevin function 
to model the dependence on current $I$ and a 
hyperbolic function for the dependence on height $z$,
\begin{align}
   B_z(z, I) = a\left(\coth(b_I I) - \frac{1}{b_I I}\right)
  \left(\frac{a_z}{z- b_z} + c_z\right).
\end{align}
with fit parameters $a = 4.647$, $b_I = 0.332\,\mathrm{A^{-1}}$, 
$a_z = 286.7\cdot 10^{-6}\,\mathrm{Tm}$, 
$b_z = -1.104\cdot10^{-3}\,\mathrm{m}$, and 
$c_z = 10.86\cdot10^{-3}\,\mathrm{T}$. 
A plot of the fit curve 
together with the measured magnetic field is shown in 
the Supporting Information. 
The fit describes the magnetic field with an error below 
1 \% in the neighborhood of the capsule.

\subsection{Magnetization curve}
\label{sec:mag}

The magnetization curve of the ferrofluid is calculated from the particle
size distribution of the magnetite nanoparticles. 
 The particle size distribution
is obtained from dynamic light scattering (see Supporting Information)
and given in Table
\ref{tab:diameters}, where the relative
frequency $n_i$ of 
 particle diameters $d_i=5-14\,\mathrm{nm}$ are given. 
 The particle size distribution has a 
maximum around $d = 7-8\,\mathrm{nm}$.

\begin{table}
  \begin{tabular}{cc||cc}
	$d_i$ [nm] & $n_i$ [\%] & $d_i$ [nm] & $n_i$ [\%]\\
	\hline
	5 & 2.0 & 10 & 8.6\\
	6 & 14.1 & 11 & 7.6\\
	7 & 23.9 & 12 & 5.5\\
	8 & 23.9 & 13 & 3.8\\
	9 & 16.1 & 14 & 2.4
   \end{tabular}
\caption{Particle size distribution: relative
frequency $n_i$ of particle diameter $d_i$.}
\label{tab:diameters}
 \end{table}

\begin{figure}[t]
  \includegraphics[width=0.48\textwidth]{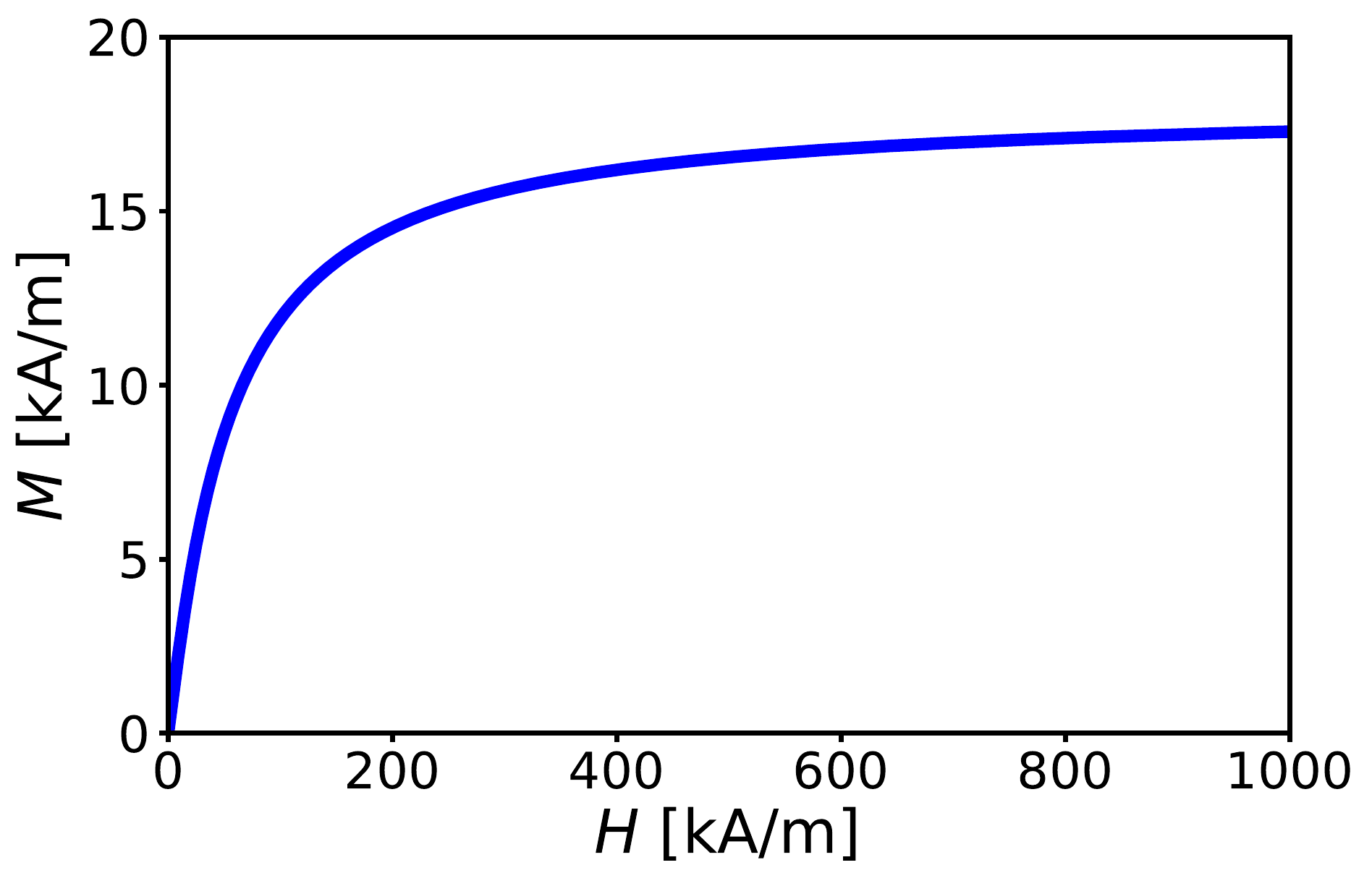}
\caption{ 
Magnetization curve of the ferrofluid used in the experiment
 according to eq \ref{eq:MagCurve}. 
}
 \label{fig:MagCurve}
\end{figure}

According to Ref.\ \citenum{Rosensweig1985}
 the magnetization $M$ as a function 
of the field strength $H$ is then given by 
\begin{align}
\label{eq:MagCurve}
\begin{split}
M(H) &= M_s\sum\limits_in_i(d_i - d_s)^3\times \\
 & L\left[\frac{\mu_0 M_d H}{k_b T}\frac{\pi}{6}(d_i-d_s)^3 \right]
  \bigg{/}\sum\limits_in_id_i^3
\end{split}
\end{align}
with the Langevin function $L(x) = \coth(x) - 1/x$. $M_s =
34650\,{\text{A}}{\text{m}^{-1}}$ is the saturation magnetization of the
magnetite particles without sterical stabilization and $M_d =
446000\,{\text{A}}{\text{m}^{-1}}$ denotes the bulk magnetization of
magnetite. The diameters of the nanoparticles are reduced by $d_s$,
because the crystal order of the outer surface layer of magnetite is disturbed
by the dispersing agent. This lowers also the effective saturation
magnetization of the ferrofluid. Following Ref.\ 
\citenum{Rosensweig1985} we use
$d_s = 1.66\,\text{nm}$. The 
resulting magnetization curve is shown in Figure
\ref{fig:MagCurve}.

\subsection{Numerical procedure}

The calculation of the magnetic field inside the capsule and the ferrofluid is
done numerically by a coupled FEM/BEM
\cite{Costabel87,Wendland88,Arnold83,Arnold85,Ligget81,Lavrova05,Lavrova06,Harris1993}.
The shape of the elastic shell and the magnetic field inside the capsule form
a coupled problem that has to be solved self-consistently: 
Changes in the shape of the elastic shell cause changes in the
 magnetic field distribution inside the capsule, which, in turn, 
changes the magnetic forces in eq \ref{eq:stress_tensor}
onto the capsule shell and
thus the capsule shape. Therefore, an iterative
solution scheme has to be used. For a given trial 
shape, we calculate the magnetic
field inside the capsule using only the externally applied field. 
 Using the new field,
we can recalculate the magnetic forces and a new equilibrium shape.
This is iterated until shape
and field converge to a fixed point. This iterative solution scheme for a
ferrofluid-filled capsule has been introduced and is explained in detail in
Ref.\ \citenum{Wischnewski2018}.

\section{Experiments and Results}

\subsection{Mechanical characterization of the capsule shell}

We first characterize the 
capsules mechanically and determine the two-dimensional Young
modulus $Y_{2\text{D}}$ and the surface Poisson ratio $\nu$. These values are 
subsequently used in the theoretical model for magnetic deformation.
We determine
 mechanical parameters of the shell with two independent methods,
compression between parallel plates and spinning capsule deformation. 
If the Poisson's ratio $\nu$ is known
both methods give the two-dimensional elastic
 modulus $Y_{2\text{D}}$. A separate measurement 
 of $\nu$ is, however, difficult and often a generic value 
is assumed
 \cite{ben_messaoud_influence_2016,leick_deformation_2010}. 
By combining the results of the two independent measurements,
we overcome this problem and 
 obtain two independent 
equations for the two unknowns $Y_{2\text{D}}$ and $\nu$,
which enables us to determine both quantities \cite{leick_rheologische_2011}.

The first method, the compression between
parallel plates (capsule squeezing) is a well-known technique for the
mechanical characterization of capsules
\cite{carin_compression_2003,fery_mechanical_2007}. 
A capsule is placed between two
parallel plates and the force is measured as a function of the
displacement. 
From the force-displacement compression curve, 
elastic moduli 
of the capsule shell can be calculated. 
There are different methods to analyze the force-displacement curves.
Some methods, like the one developed by Barth\`{e}s-Biesel
 \textit{et al.}, use a fit of the whole curve
\cite{carin_compression_2003,fery_mechanical_2007}. As shown in an
earlier publication water leaks out of alginate capsules during
compression \cite{degen_magneto-responsive_2015}. Therefore,
we only analyze the
linear regime with low forces in order to minimize the influence of 
water leakage. 

To describe the compression of a capsule between parallel 
plates in the initial linear regime,
the model of Reissner for a point force acting 
on the apex of the capsule can be used
 \cite{fery_mechanical_2007,reissner_stresses_1946,reissner_stresses_1946-1}.
 In this model the force $F$ depends linearly on the displacement $d$ as
\begin{equation}
 F = \frac{4 Y_{2\text{D}} h}{R_0 \sqrt{3 (1-\nu^2)}} \: d.
 \label{eq:reissner}
\end{equation}
Reissner's linear theory applies to an unpressurized shell with 
 purely elastic tensions and rest radius $R_0$.
Here, we need to generalize this model to include the interfacial
 tension $\gamma$, which generates a stretching tension in the 
shell already before indentation by the force.
We find that the effect of an additional interfacial tension $\gamma$ 
is equivalent to the effect of an internal pressure $p_0=2\gamma/R_0$
given by the Laplace-Young equation; such pressurized capsules 
have been studied previously by 
Vella {\it et al.} in Ref.\ \citenum{Vella2012}.
Using this equivalence 
we obtain a modified linear force-displacement 
relation 
\begin{align}
 F = \frac{4 Y_{2\text{D}} h}{R_0 \sqrt{3 (1-\nu^2)}} \: G(\tau)\:d
 \label{eq:reissner_modified}
\end{align}
with
\begin{align}
 G(\tau) &=
 \frac{\pi}{2}\frac{(\tau^2-1)^{1/2}}{\operatorname{artanh}{(1-\tau^{-2})^{1/2}}}
 \\
 \tau &= 3(1-\nu^2)\left(\gamma/Y_{2\mathrm{D}} \right)^2\left(R_0/h \right)^2
\nonumber
\end{align}
(more details on the derivation are given in the Supporting Information). 
Note that a vanishing interface tension ($\gamma = 0$) leads to 
$G(0) = 1$, and we recover the Reissner result (eq \ref{eq:reissner}).
 To obtain the elastic moduli from this linear model, 
 also the radius of the undeformed capsule $R_0$ and the
 thickness of the shell $h$ are needed.

The second method for the determination of $Y_{2\text{D}}$ and $\nu$ is the
spinning capsule method. Originally, the spinning drop method was developed
for interfacial tension measurements
\cite{vonnegut_rotating_1942,princen_measurement_1967}. A drop is
placed in a capillary filled with a liquid of higher density $\rho$.
 During rotation of the capillary the drop deforms; the shape of the
deformed drop depends on the rotation frequency and 
interfacial tension $\gamma$. For the deformation of liquid-filled capsules,
Pieper \textit{et al.} developed a
model that allows one to obtain in an analogous fashion 2D
elastic moduli from the capsule deformation \cite{pieper_deformation_1998}.
A sketch of the method is shown in the Supporting Information.
Initially (at small rotation speed), 
 the capsule is in an undeformed (quiescent) state. Then
capsule deformation is measured as a function of the angular rotation
frequency $\omega$. The capsule deformation is quantified by the 
Taylor deformation $D$, 
\begin{equation}
 D = \frac{l-b}{l+b},
 \label{eq:Taylordef}
\end{equation}
which is determined from the length $l$ and the
width $b$ of the capsule.
 For deformations within the linearly elastic regime, 
the Taylor deformation depends on 
the angular rotation frequency $\omega$ via 
 \cite{pieper_deformation_1998}
\begin{equation}
 D = -\Delta \rho \omega^2 R_0^3 \frac{(5 + \nu)}{16 Y_{2\text{D}}},
 \label{eq:SD_original}
\end{equation}
 where $R_0$ is the radius of the undeformed capsule,
  $\Delta \rho$ is the density difference between the inside and 
outside liquid phases.
 Equation \ref{eq:SD_original}
 is only valid for shells with purely elastic tensions 
 and without
 an interfacial tension. Again we need to generalize the theory 
 to include an interfacial tension $\gamma$. 
This can be done by observing that 
 the linear response of the 
capsule deformation in spinning drop experiments 
 is actually equivalent to the linear
 deformation response of a ferrofluid-filled capsule in a uniform external
 magnetic field \cite{Wischnewski2018} 
 with a magnetic susceptibility $\chi=-1$ and a magnetic field strength 
 $\mu_0H^2 = \Delta\rho R_0^2\omega^2$.
This equivalence arises 
because 
 both centrifugal forces exerted by the interior liquid 
 in the spinning capsule geometry and 
 magnetic forces exerted by the ferrofluid in an external field 
 are normal surface forces. Moreover, for a spherical shape 
 they have the 
same position-dependence for 
 $\chi=-1$ and the same magnitude 
 if we set $\mu_0H^2 = \Delta\rho R_0^2\omega^2$.
 In Ref.\ \citenum{Wischnewski2018} the deformation of a 
ferrofluid-filled magnetic capsule has already been considered 
also in the presence of an interfacial tension $\gamma$, and we 
can exploit 
the equivalence of both problems and 
 adapt the results of Ref.\ \citenum{Wischnewski2018}
for the linear response (more details are 
given in the Supporting Information). We find
 \begin{align}
   D = -\Delta \rho \omega^2 R_0^3 
 \frac{(5 + \nu)}{16 [Y_{2\text{D}} + (5+\nu)\gamma] },
  \label{eq:SD}
 \end{align}
that is, in the presence of an interfacial tension we simply have to 
replace $Y_{2\text{D}}$ in eq \ref{eq:SD_original} by 
$Y_{2\text{D}} + (5+\nu)\gamma$.

The two unknown material parameters 
Young's modulus $Y_{2\text{D}}$ and surface Poisson ratio $\nu$ can now be
obtained from solving the two eqs 
 \ref{eq:reissner_modified} and \ref{eq:SD} simultaneously, 
which can only be done numerically for $\gamma\neq 0$ in general.
In order to do so, also the rest radius $R_0$ and the shell
thickness $h$ of the spherical capsules are needed. 
The mean capsule radius was determined from image analysis (see Materials 
and Methods) as 
 794 $\pm$ 87$\,\mathrm{\mu m}$. 
The shell thickness was determined by SEM (see Materials and Methods 
and Supporting Information). 
The shell thickness estimated by SEM was under $1 {\rm \mu mu}$. 
As SEM is performed in
ultra-high vacuum, swelling has to be taken into account.
Because we were not able to resolve the hydrated capsule shell
by optical microscopy, its thickness has to be
below $5\,\mathrm{\mu m}$.
We conclude that the shell thickness of hydrated capsules is
approximately $3\,\mathrm{\mu m}$ but with a relatively 
high error around $1\,\mathrm{\mu m}$.

\begin{figure}[h]
\centering
\includegraphics[width=.48\textwidth]{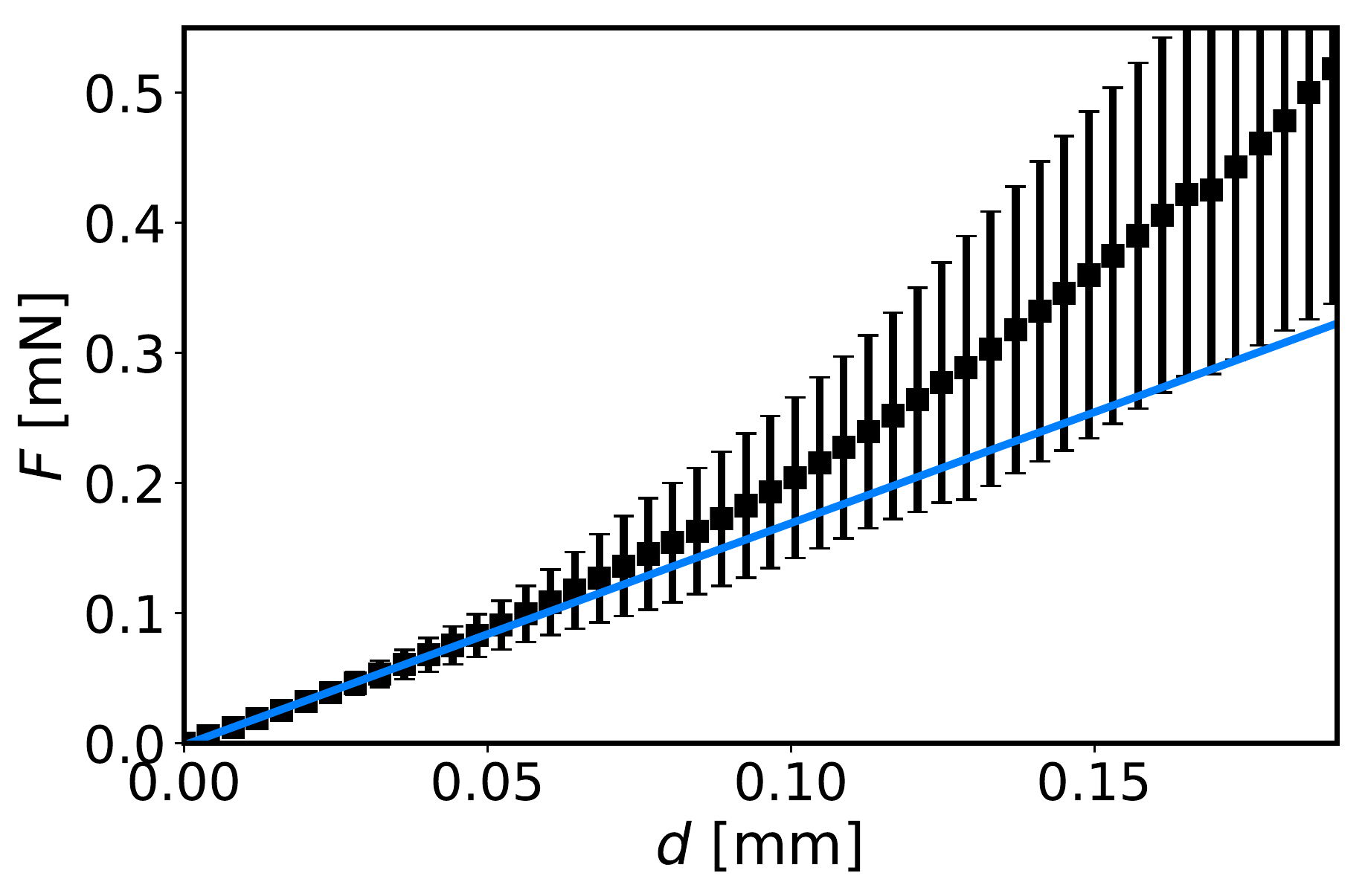}
\caption{ 
 Averaged force($F$)-displacement($d$) curve for capsule compression 
between parallel plates and fit to the linear regime.
}
 \label{fig:ErgCC}
\end{figure}

Figure \ref{fig:ErgCC} shows the results for the
force-displacement curves in the capsule compression tests. 
We measured 15 individual capsules,
 and the respective force values for the
displacement were averaged to gain information on replicability. 
 In Figure
\ref{fig:ErgCC} the averaged force-displacement curve is shown.
The linear regime can be seen clearly. In this regime the error bars are also
relatively low. Higher error bars with increasing compression are an
effect of capsule size variations. For the calculation of the
elastic moduli, the analysis was performed for each capsule separately with
the respective radii and an average value $FR_0/d$ is extracted 
from the linear regime (blue curve in Figure \ref{fig:ErgCC}).

\begin{figure}[h]
 \centering
 \includegraphics[width=.48\textwidth]{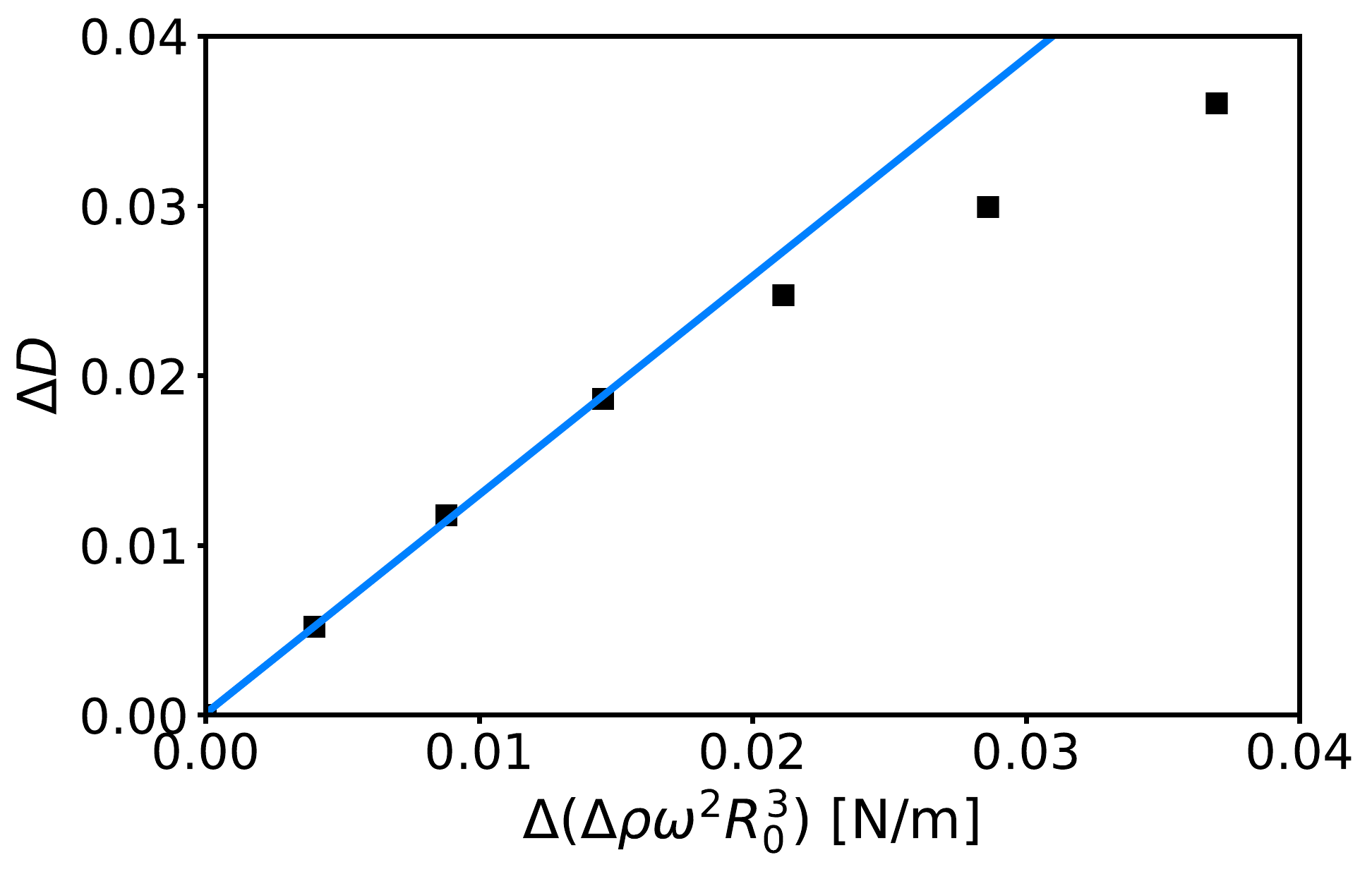}
 \caption{
 Results of an exemplary spinning capsule experiment for the 
 difference in 
 Taylor deformation $\Delta D$ (see eq \ref{eq:Taylordef}) 
 between deformed and initial state 
 as a function of $\Delta (\Delta \rho \omega^2 R_0^3)$.
}
 \label{fig:ErgSD}
\end{figure}

Figure \ref{fig:ErgSD} shows the results of the spinning capsule 
experiments for one exemplary capsule. 
Because of small mechanical forces which act during the capsule synthesis, we
 observed small initial Taylor deformation values of $D_i = 1-2\%$, 
which were not caused by deformation in the centrifugal field. 
Also the initial quiescent capsule state is measured at $\omega_i>0$. 
Therefore, we subtract the initial deformation and initial 
angular frequency and use the 
Taylor deformation difference $\Delta D=D-D_i$ and the 
difference $\Delta \omega = \omega-\omega_i$ 
for the analysis. 
Figure \ref{fig:ErgSD} shows the expected linear dependence of 
$\Delta D$ from the parameter $\Delta (\Delta \rho \omega^2 R_0^3)$ 
according to eq \ref{eq:SD}. The deviations at higher rotation frequencies 
are caused by water leakage from the capsule, which was also 
observed in previous studies \cite{zwar_production_2018}. 
 For the calculation of the
elastic moduli, the slope $\Delta D /\Delta (\Delta \rho \omega^2 R_0^3)$ 
is determined 
from the initial linear regime (blue curve in Figure \ref{fig:ErgSD}).

 The 
results from both capsule deformation methods are used to calculate the
surface Poisson ratio $\nu$ and the 2D Young modulus $Y_{2\text{D}}$
by solving eqs \ref{eq:reissner_modified} and \ref{eq:SD}
simultaneously. 
We find 
\begin{align}
Y_{2\text{D}} &= 0.186 \pm 0.040\,\mathrm{N/m}\\
\nonumber
&\text{and} \\
1 - \nu &= (1.9\pm 2.8) \cdot 10^{-7}.
\label{eq:Ynu}
\end{align} 
Because $\nu$ is extremely close to unity, the two eqs
\ref{eq:reissner_modified} and \ref{eq:SD} 
can be decoupled with negligible error,
which leads to the following simplification:
We set $\nu = 1$ in eq \ref{eq:SD} 
and calculate $Y_{2\mathrm{D}}$ only from the spinning capsule experiment.
This value is then used to calculate $\nu$ via eq \ref{eq:reissner_modified}.
This decoupling has the additional benefit
 that we can determine $Y_{2\mathrm{D}}$ from eq \ref{eq:SD} 
for the spinning capsule experiment
effectively independently of the shell thickness $h$, 
which can only be measured with 
a relatively high error as explained above.

As compared to other
calcium alginate capsules, the moduli are very low
\cite{ben_messaoud_influence_2016,zwar_production_2018}. This is likely an
effect of the encapsulated components, that is, the nanoparticles or the
specific oils used. In emulsion encapsulation, the moduli were
lowered by the addition of nanoparticles
\cite{degen_magneto-responsive_2015}. As the main goal was to obtain easily
deformable capsules, this was achieved by creating very thin shells. A
remarkable result is the high surface Poisson ratio, which points to an
area-incompressible shell \cite{pieper_deformation_1998}. 
Therefore, we further check
this finding by adjusting this parameter in 
 the numerical calculation and analysis of the magnetic deformation of 
the capsule shape to fit the experimental shapes.
For Young's modulus we use the measured value 
 $Y_{2\text{D}} = 0.186\,\text{N}/\text{m}$
 in the numerical calculations 
(see Table \ref{tab:Parameters}).

\begin{figure}[t]
 \includegraphics[width=82.5mm]{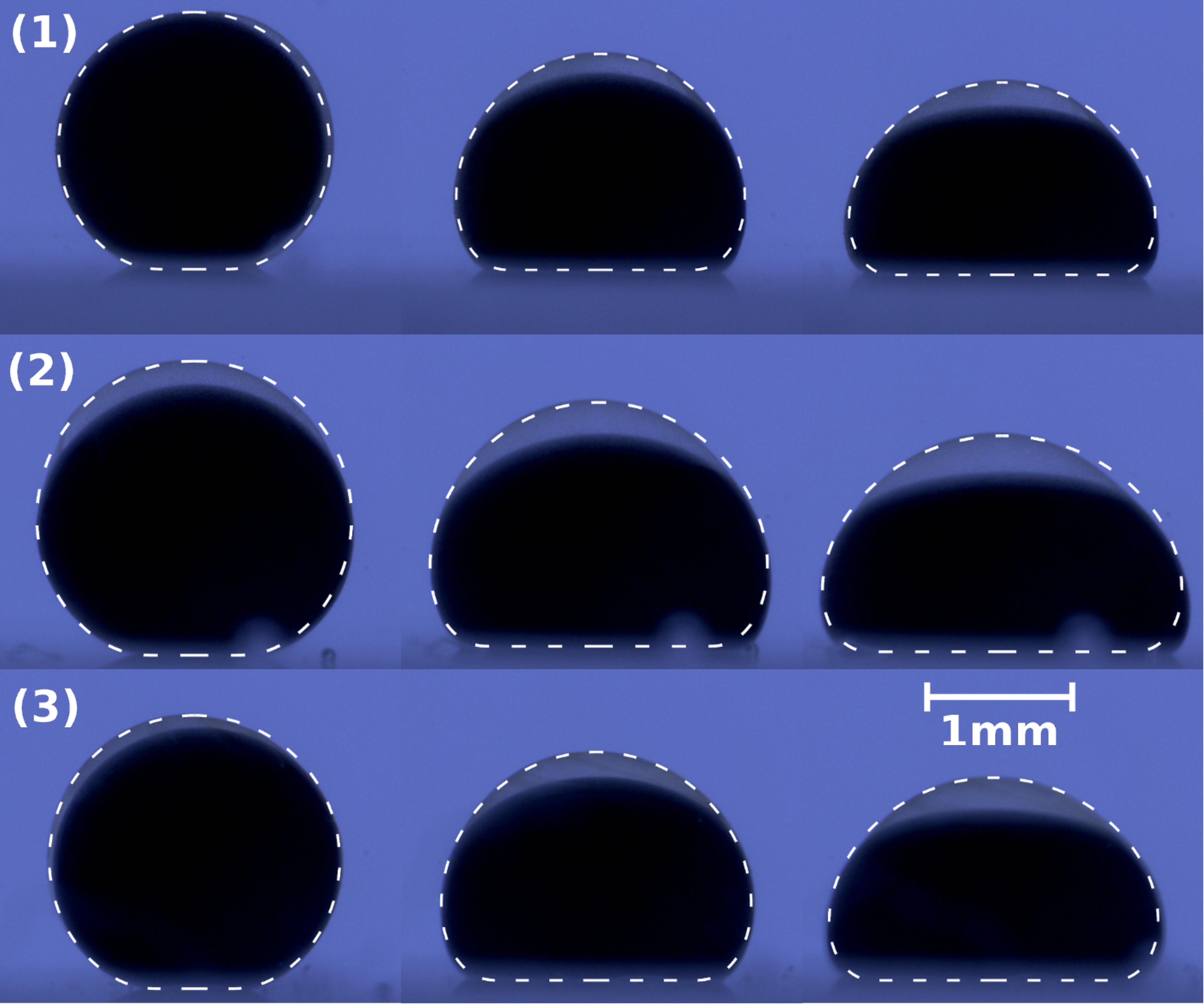}
 \caption{ 
 Photographic images of of three different capsules with 
  radii (1) $R_0=0.903\,\text{mm}$, (2) $R_0 = 1.044\,\text{mm}$, 
 and (3) $R_0 = 0.965\,\text{mm}$, each for increasing 
 magnetic fields with 
  $I=0\,\text{A}$ (left), $I=2\,\text{A}$ (middle) and $I=5\,\text{A}$
  (right). The dashed lines describe numerically
  calculated capsule contours using the parameters in 
 Table \ref{tab:Parameters}. A surface Poisson ratio $\nu = 0.946$ gives 
 the best fit to the experimental shapes. 
}
  \label{fig:Contour123}
\end{figure}

\subsection{Magnetic deformation}

We prepared different capsules and deformed them with the previously described
inhomogenous field by increasing the electric current in the coil up to
$5\,\text{A}$. This corresponds to magnetic field 
variations up to $50\,\mathrm{mT}$ over the size of the capsule.
Photos of three exemplary
capsules that have reached their steady state of deformation 
 are shown in Figure \ref{fig:Contour123}. 
In Figure \ref{fig:Contour123} 
we compare these photos with numerically calculated contours (with dashed
lines), which were 
generated with the parameter values 
in Table \ref{tab:Parameters}. These are the experimentally measured 
parameter values except for the surface Poisson ratio $\nu$. In particular, 
we consider the mechanical measurements of the Young modulus 
$Y_{2\text{D}} = 0.186\,\mathrm{N/m}$
 to be exact and employ this value in the numerics. 
The Poisson ratio $\nu = 1- 1.9 \cdot 10^{-7}$ from mechanical characterization 
 is extremely close to unity; 
it depends very sensitively
on the shell thickness $h$, which is not easy to determine reliably. 
Therefore, we regarded $\nu$ as a free parameter and 
adjusted $\nu$ for the best fit 
to the experimental shapes, which gives a slightly lower value 
$\nu \approx 0.946$.
This confirms that the surface Poisson ratio of the shell is close to unity.

\begin{figure}
  \includegraphics[width=1.0\linewidth]{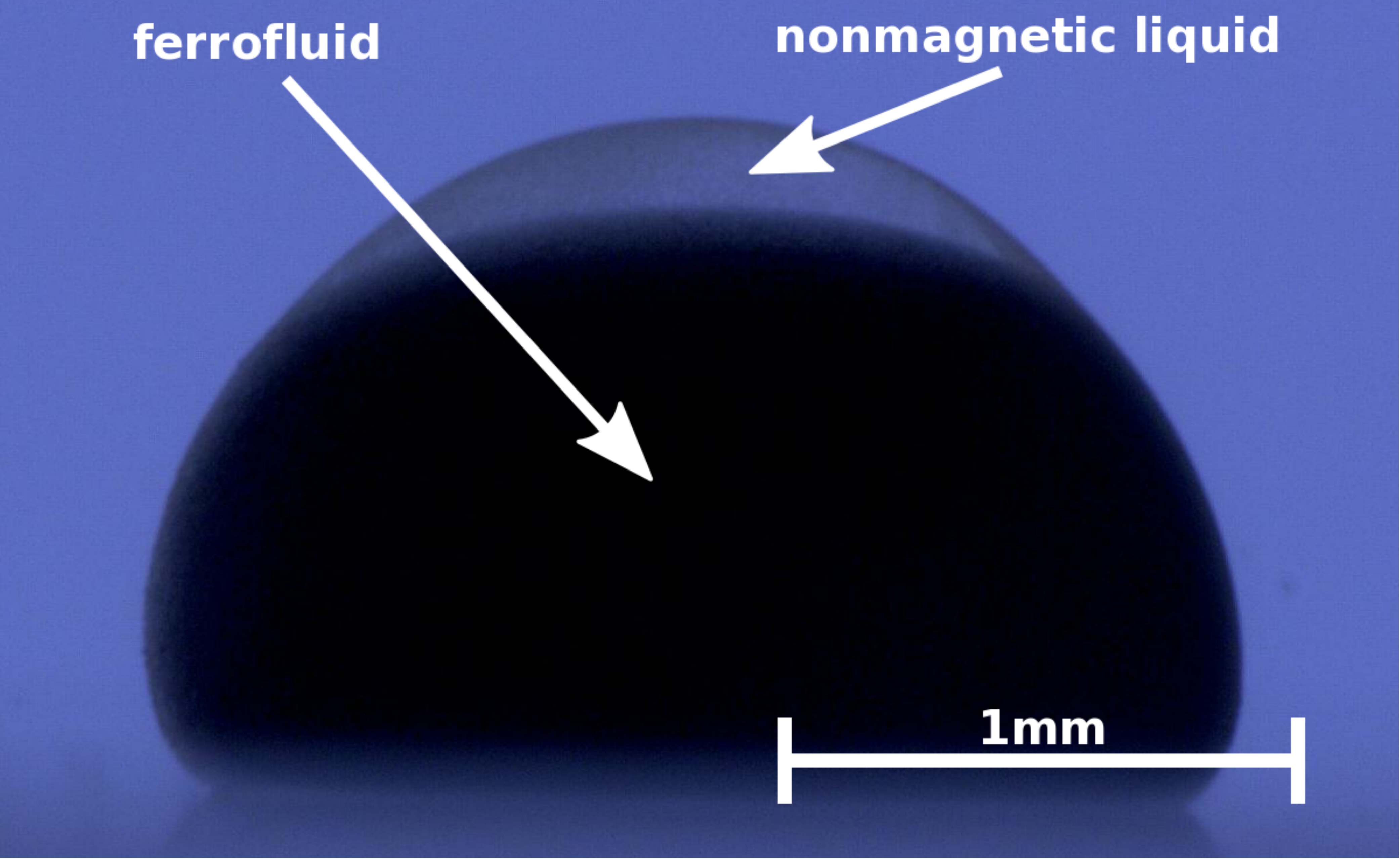}
 \caption{ 
 Magnified photographic image
 of capsule 1 for $I=5\,\text{A}$. A
  non-magnetic transparent liquid,
 which is clearly phase-separated from the ferrofluid, 
 is pushed to the upper part of the capsule. 
}
   \label{fig:capsule1_phases}
\end{figure}

 There are minor deviations between
experimental capsule shapes and the calculated contours, which 
 are probably caused by inhomogeneities and
asymmetries in the shell thickness.
 In addition, we also observe in
 Figure \ref{fig:Contour123} that a small amount of transparent non-magnetic
liquid was also caught inside the capsules. As the
 transparent phase spreads lense-like at the surface of the ferrofluid
 droplet, we assume that this liquid is polar and likely contains water
 that diffused through the membrane. This liquid is pushed to the upper
part of the capsule when the ferrofluid is pushed downwards 
by the inhomogeneous magnetic field as can be seen in the deformed
capsules in Figure \ref{fig:Contour123} and in the close-up in Figure
\ref{fig:capsule1_phases}.  Because
this fluid does not contain magnetic nanoparticles, it accumulates at the top
of the capsule and, thus, at the place with the lowest field strength during
deformation.
The appearance of this liquid could not be
prevented. The additional error in comparison to the numerical model
caused by that liquid should be small, however, because
the magnetic stress from the ferrofluid
acts on the interface between the ferrofluid and the non-magnetic liquid and is
transmitted to the elastic shell by the liquid. Therefore, the behaviour of
the whole capsule should be comparable to a capsule with only a ferrofluid
inside.

\begin{figure}
  \includegraphics[width=82.5mm]{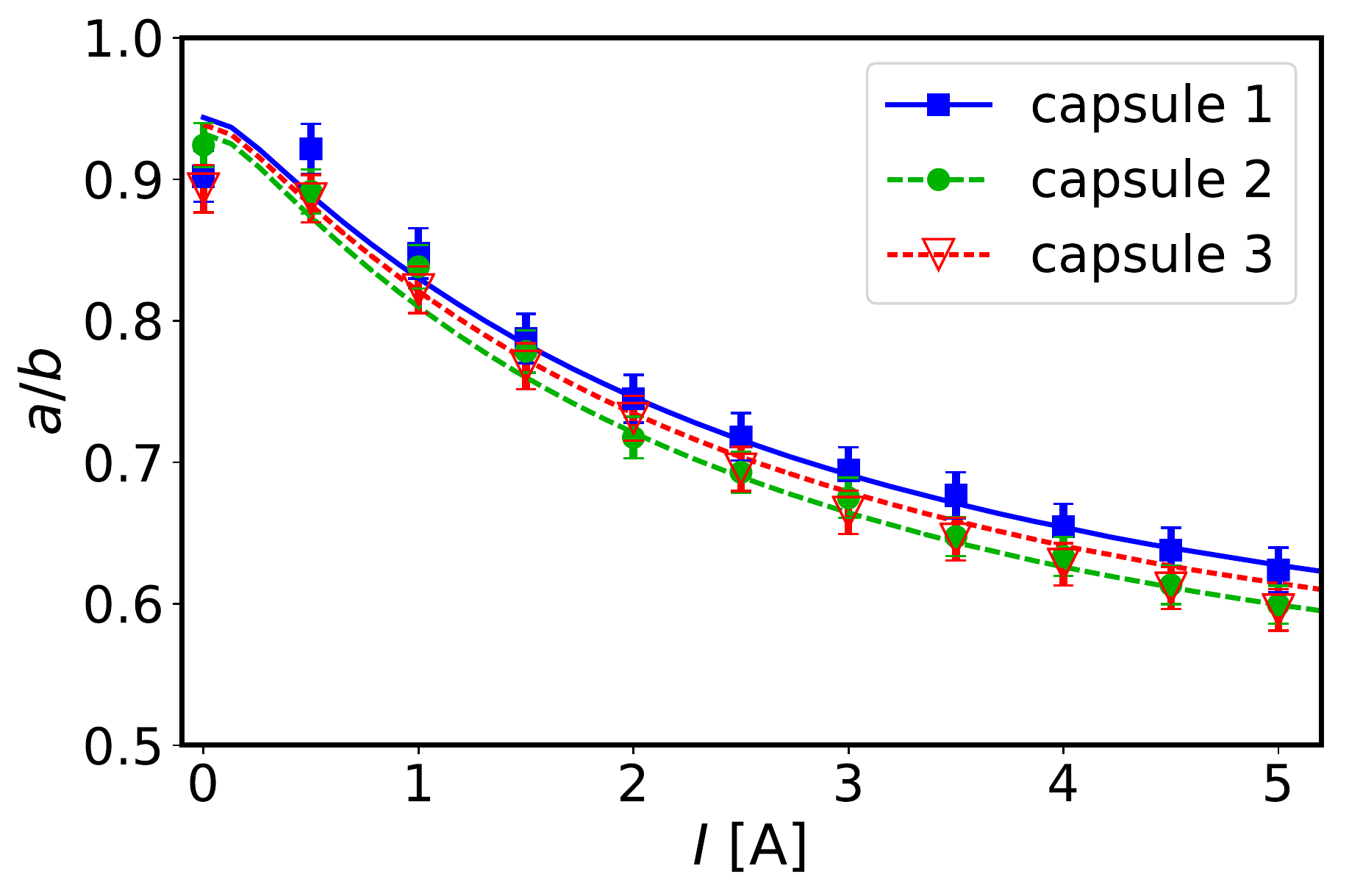}
  \caption{ 
  Ratio of capsule height $a$ to width $b$ for
   increasing current $I$. Markers denote capsules from the experiment,
   while solid lines represent numerical data generated with 
  the best fit $\nu \approx 0.946$. The upper curve (blue,
   squares) are data for 
   capsule 1 from Figure \ref{fig:Contour123}, the middle curve
   (red, triangles) and lower curve (green,
   circles) data for capsules 3 and 2, respectively. 
}
  \label{fig:ab123}
\end{figure}

In order to perform a quantitative comparison between numerics and the
experiment, we measure the ratio of the capsule's height ($a$) to the maximum
width ($b$), measured parallel to the plate below the capsule. The results of
these measurements are shown in Figure \ref{fig:ab123}. 
We achieve the best agreement between 
numerical calculations and experimental data for $\nu \approx 0.946$.
 As also the 
comparison of capsule shapes in 
Figure \ref{fig:Contour123} indicates, there is a good agreement between 
numerical calculations and the experiment for $\nu \approx 0.946$. 
The only point with some higher
deviation is at $I = 0$, that is, in the absence of 
 any magnetic field. Then the deformation
of the capsule is only caused by gravity and, thus, is in total relatively 
 small.
This makes the capsule side ratio $a/b$ very prone to asymmetries of the
shell. In addition, the capsule is not a perfect sphere after 
 membrane gelation as it is assumed in the numerical calculation.

We can also check that 
the capsule's volume $V_0$ remains constant during the magnetic 
deformation process. Any loss of the inner fluid through the shell would be
 visually detectable during contact with the outer fluid. In addition,
 analysis of the capsule photos for the capsule contour $r(z)$ 
and calculation of the volume
$V_0= \int dz \pi r^2(z)$ does not indicate any loss of volume
(see Supporting Information).
We conclude that the elastic shell is impermeable 
for the inner and outer fluid
 under the employed  experimental conditions and 
 during magnetic deformation.

\begin{figure}
\centering
 \includegraphics[width=.48\textwidth]{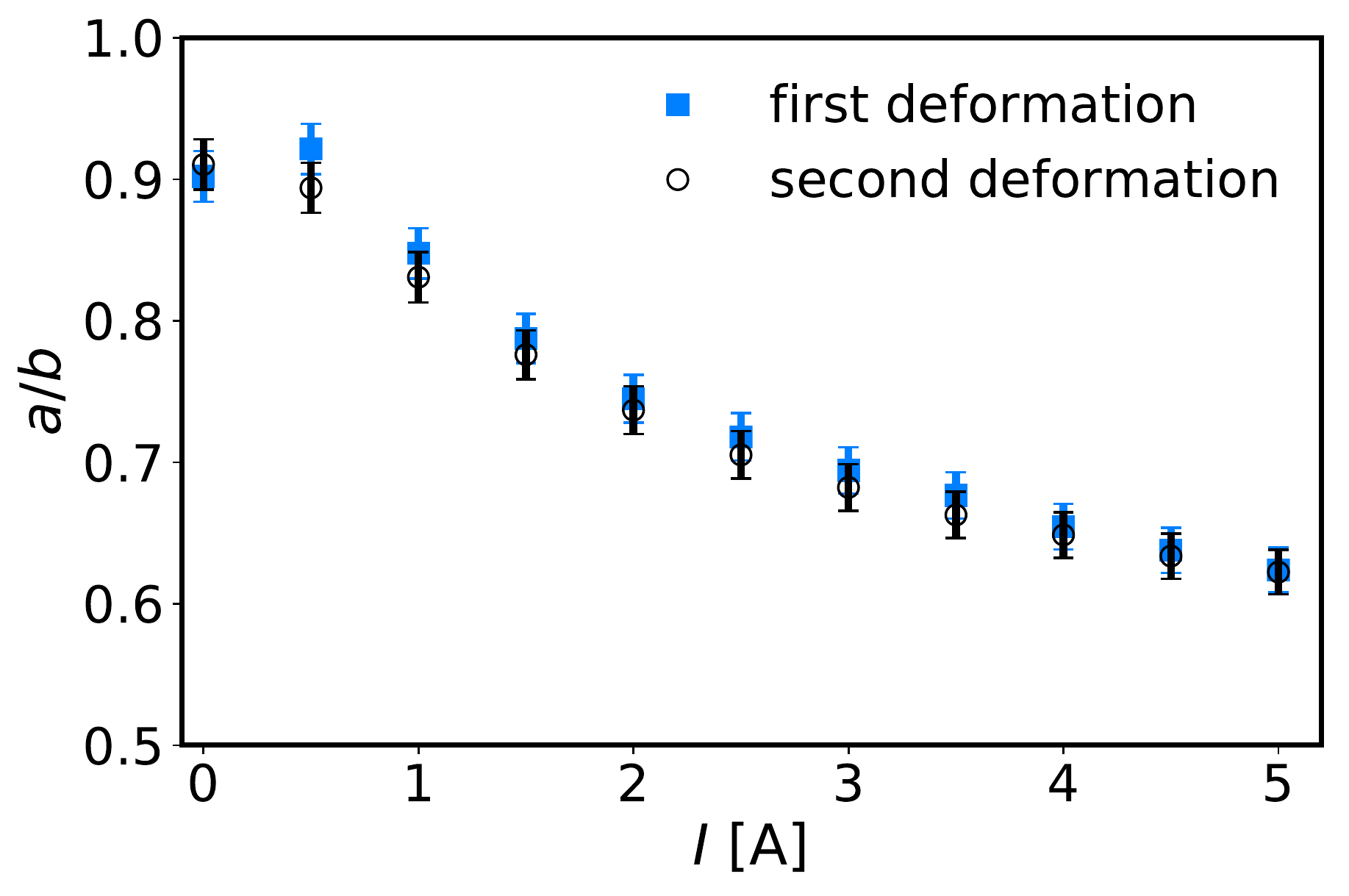}
 \caption{ 
 Ratio of the height $a$ of capsule 1 to the width $b$ for
  increasing current $I$. Squares denote the first deformation. Circles
  represent a second measurement with the same capsule after deforming and
  relaxing the capsule. }
  \label{fig:ab_reversibility}
\end{figure}

An important characteristic for possible applications is the question if
the material is weakened by the deformation, that is, whether aging 
effects occur. We checked this by performing a second deformation cycle.
 The results are shown in Figure \ref{fig:ab_reversibility}
exemplarily for the first capsule.
Second and first deformation cycles agree well which
indicates that no significant damage or aging in the elastic shell 
occurs during the deformation process.
Also the preserved volume during deformation hints at a 
reversible deformation process.

\begin{figure*}
\centering
\includegraphics[width=0.99\textwidth]{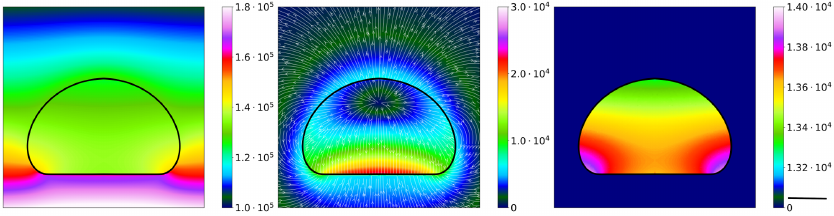}
 \caption{ 
 Numerically calculated distributions of magnetic field strength (left),
 stray field (middle), and magnetization (right) for capsule 1 
 (see Figure \ref{fig:Contour123}) at $I = 5$\,A.
 The solid line represents the capsule's elastic shell.
 Left: Color codes for the absolute value of   the total
  magnetic field $|\vec{H}|$ (in A/m). 
 Middle: Color codes for
 the absolute value of the capsule's own magnetic stray field 
  $\vec{H}_{\text{stray}}
  = \vec{H} - \vec{H_{\text{extern}}}$ (in A/m), arrows indicate the field
  direction. 
  Right: Color codes for the absolute value of the magnetization 
  $|\vec{M}|$ (in A/m). Note that magnetization variations 
 are very small. 
}
  \label{fig:field_plot}
\end{figure*}

Having established a good agreement between experiment and theory as evidenced
by Figures \ref{fig:Contour123} and \ref{fig:ab123}, we can use the
theoretical model to access important physical quantities that are
experimentally hardly accessible, such as the detailed distributions of
magnetic field and stresses. The distribution of the magnetic field inside
the deformed capsules is easy to calculate from our model. 
 Figure \ref{fig:field_plot}
shows the magnetic field inside and outside of capsule 1 at $I = 5$\,A, 
the stray field generated by the 
magnetized capsule, and the magnetization inside the capsule. The external
magnetic field has a high gradient, but the total field inside the capsule is
surprisingly homogenous. The magnetization of the ferrofluid decreases the
total field in the lower section of the capsule, whereas it increases the total
field in the upper part because it counteracts the external magnetic field 
according to Lenz's law. This results in smaller field gradients
inside the capsule. Magnetization variations inside the 
capsule are small.

\begin{figure*}
\centering
 \includegraphics[width=1.0\textwidth]{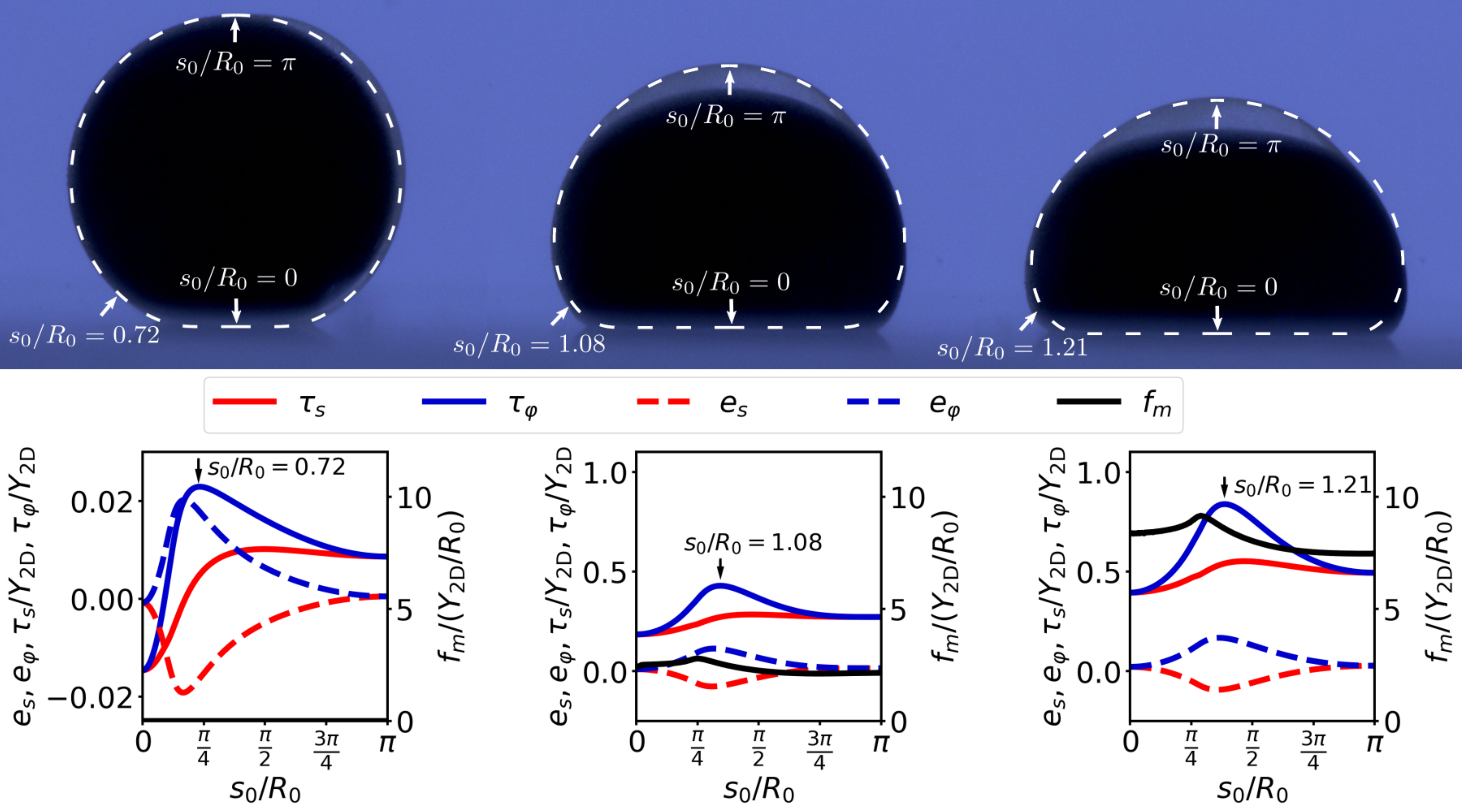}
 \caption{ 
 Stresses and strain distributions for different deformation states
  of capsule 1 (see Figure \ref{fig:Contour123}) for 
  zero magnetic field (left),
  $I = 2$\,A (middle) and $I = 5$\,A (right). The left scales 
  show the strain in meridional ($e_s$) and circumferential
  ($e_\varphi$) direction and the corresponding elastic stresses ($\tau_s$,
  $\tau_\varphi$ in units of $Y_{2\mathrm{D}}$) inside the shell. The right
  scale shows the magnetic pressure $f_m$ (in units of $Y_{2\mathrm{D}}/
  R_0$). The corresponding positions on the surface are given by the
  arc length $s_0$ of the undeformed reference sphere (in units of $R_0$,
   $s_0 \in [0,\pi]$). 
 For comparison, the interface tension has a 
 value $\gamma = 0.054\,Y_{2\mathrm{D}}$. This shows that the interfacial 
  tension $\gamma$ 
 is the dominating stress for small deformations ($e_{s,\varphi}<0.05$) 
  by gravity 
 and at small magnetic field, while the strongly 
 deformed shell is completely dominated by elasticity.
}
  \label{fig:stresses}
\end{figure*}

Also the elastic stress state inside the elastic shell as given by 
the stress
and strain distribution becomes accessible by numerical calculation. Figure
\ref{fig:stresses} shows the force/stress/strain distribution along the
capsule contour for capsule 1 in different stages of deformation. The position
of the highest elastic stress is marked, which is at the 
highly curved region at the bottom of the capsule. We can infer that 
 the capsule will most likely rupture at this position
 if the magnetic field is further increased.

 \subsection{Poisson's ratio $\nu$}
  \label{sec:poisson}

For the numerics, we consider the measurement of the Young-Modulus to be exact
with a value of $Y_{2\text{D}} = 0.186\,\mathrm{N/m}$. Varying the surface
Poisson ratio $\nu$, we find the best match between numerics and experiment
for $\nu \approx 0.946$. 
This confirms a surface Poisson ratio of the shell
very close to unity and is only slightly lower than the value $\nu = 1 -
1.9\cdot10^{-7}$ from mechanical characterization. 
This represents an
explicit measurement of the surface Poisson ratio for an alginate membrane.
Other measurements of the surface Poisson ratio of alginate shells are
only available for barely comparable systems and solely rely 
on mechanical characterization
\cite{leick_rheologische_2011, zwar_production_2018};
in the literature often a generic value (such as $\nu=1/3$ or
$\nu\approx0.5$) is assumed
\cite{ben_messaoud_influence_2016,leick_deformation_2010}.

The two-dimensional 
area compression modulus $K_{2\text{D}} = Y_{2\text{D}}/2(1-\nu)$ 
diverges for $\nu$ close to unity. 
Therefore, surface 
 Poisson ratios close to unity are remarkable as they 
imply a polymerized shell that is
 nearly area preserving during deformation.
For Poisson ratios $\nu$ close to unity, strong elastic
 deformations also become quite sensitive to changes of $\nu$, as 
 the diverging compression modulus also gives rise to 
diverging elastic stresses containing factors $1/(1-\nu)$. 
 This allows us to determine $\nu$ rather precisely by 
fitting numerical to experimental shapes 
 (see also Supporting Information).

In a three-dimensional material volume-incompressibility 
during deformation 
corresponds to a three-dimensional Poisson ratio 
$\nu_{3\text{D}}=0.5$ (the three-dimensional compression 
modulus is $K_{3\text{D}} = Y_{3\text{D}}/3(1-2\nu_{3\text{D}})$). 
Three-dimensional Poisson ratios close 
to $\nu_{3\text{D}}=0.5$ are common for three-dimensional 
polymeric materials, where volume-incompressibility is, for example, 
explicitly implemented 
in the Mooney-Rivlin constitutive relation commonly used 
for polymeric materials \cite{Barthes-Biesel2002}. 
For polymeric materials incompressibility can be attributed to the 
effectively incompressible densely packed liquid of monomers. 
Also alginate gels are typically regarded as volume-incompressible 
materials as confirmed by measurements, for example, 
in Refs.\ \citenum{Espona-Noguera2018,salsac2009}.
A material that is volume-incompressible in three dimensions 
($\nu_{3\text{D}}=0.5$)
forms an area-incompressible thin shell if the 
shell has constant thickness during deformation. If the thickness 
can adapt, on the other hand, we expect $\nu=\nu_{3\text{D}}=0.5$ and 
the thickness grows (shrinks) if the shell area is compressed (expanded). 
Therefore, our findings are consistent with an alginate shell consisting 
of a volume-incompressible material that maintains 
a shell of constant thickness during deformation. 
It remains to be clarified whether the inherently anisotropic structure 
of the alginate gel which results from the ionotropic gelation process
\cite{Maki2011} can contribute to the observed high surface 
Poisson ratios. 
From a chemical point of view, we expect the Poisson number 
to depend on the cross-linking and swelling degree 
of the encapsulating alginate membranes.

\section{Conclusions}

Ferrofluids are an interesting tool for different applications
involving the actuation of a fluid by magnetic fields. The
encapsulation of the ferrofluid can prevent the interaction of the fluid with
its environment. In this work we presented a novel method to encapsulate a
ferrofluid drop with a very thin elastic shell.

Using this method, a direct encapsulation of oils in alginate gels could be
achieved. We used 1-hexanol as an additive to directly dissolve
$\mathrm{CaCl_2}$ for ionotropic gelation of sodium alginate 
in an organic solvent. The magnetic nanoparticles remained
stable throughout the whole process and formed a stable ferrofluid. The
produced elastic capsules showed low Young moduli, especially compared to
other calcium alginate gels. This enables a strong deformation in relatively
weak magnetic fields, which opens possibilities for 
 applications as, for example,
switches or valves in confined spaces like microfluidic devices. In these
applications, it is essential to protect the components from contact with the
ferrofluid, as it is likely to cause unwanted reactions due to its high
reactivity. 
 Concerning a transfer to the biological or medical sector as
 motion-controllable and deformable capsule systems, applications in
 micromanipulation are imaginable. 
For any sort of industrial or medical application, a change
 of the system is necessary as chloroform is a highly volatile and toxic
 compound. This is, however,  easily achieved  as 
the oil is exchangeable and chloroform was
 only used because of its high density. 
It is also likely that the 1-hexanol can
 be substituted by other alcohols.

We first characterized the capsules mechanically using a combination 
of compression and spinning capsule techniques. We produced nearly
spherical capsules with radii about $R_0 = 1\,$mm and
mechanical characterization showed 
 a two-dimensional Young modulus of $Y_{2\text{D}}=0.186\,$N/m.
The surface Poisson ratio was close to unity. 

Then we performed
magnetic deformation in an inhomogeneous field in front of a hard constraining
wall. Our results in Figure 
 \ref{fig:ab123} demonstrate that high deformations 
with height to width ratios as low as $0.6$ could be achieved
 in inhomogeneous magnetic fields which vary by 
$50\,\mathrm{mT}$ over the size of the capsule. 
Maximal strains of about 17\% occur in the capsule shell during deformation
(see Figure \ref{fig:stresses}). 
The volume inside the capsules was
constant during magnetic deformation, that is, 
the alginate shells were impermeable. The inclusion of a
nonmagnetic, transparent, water-based liquid inside every capsule was
observed. Moreover, magnetic deformation was shown to be 
completely reversible over several deformation cycles 
without plastic or aging effects.

We presented a theoretical model and a numerical method to predict the
deformation behavior in magnetic fields by numerical calculations using 
parameter values from the experimental characterization as input parameters. 
The comparison of the capsule deformation with numerical calculations based
on a nonlinear small strain shell theory showed a good agreement between
theory and experiment (see Figures \ref{fig:Contour123} and \ref{fig:ab123}). 
In the numerical calculations we used the surface Poisson ratio $\nu$, 
which is notoriously hard to measure experimentally, as a fit parameter,
which was adjusted to optimally fit experimental shapes. 
In agreement with the mechanical characterization we obtained 
values $\nu$ close to unity ($\nu = 0.946$). 
Therefore, our results constitute the first reliable measurement 
of the surface Poisson ratio of alginate gel capsule shells. 
Surface Poisson ratios close to one suggest that the 
alginate shell deforms nearly area-preserving.
The molecular reasons for this behavior remain to be clarified.

Having established the agreement between the theoretical model and 
experimental results, we can use the numerical results to gain 
further insight into details of the deformation behavior, which are 
not experimentally accessible. 
The numerical approach gives access, for example, 
to the complete magnetic field distribution inside and outside the 
ferrofluid-filled capsule, the exerted magnetic forces, the stress
distribution and the exact deformation state in terms of strains,
see Figures \ref{fig:field_plot} and 
\ref{fig:stresses}.

\begin{acknowledgments}
We thank Monika Meuris and the ZEMM (Zentrum f\"ur Elektronenmikroskopie und
Materialforschung), TU Dortmund for providing SEM images, Iris Henkel
(Fakult\"{a}t Bio- und Chemieingenieurwesen, Lehrstuhl Technische Chemie) for
ICP-OES measurements and Julia Kuhnt for pendant drop tensiometry
 measurements.
\end{acknowledgments}

\appendix*
\section{Supporting Information}

\renewcommand{\theequation}{S\arabic{equation}}
\setcounter{page}{1}
\renewcommand{\thepage}{S\arabic{page}}

 Additional details on shape equations, on the theory of spinning drop 
and capsule compression analysis in the presence of an interfacial 
tension $\gamma$, and on the sensitivity of capsule deformation to 
Poisson number $\nu$. 
Details on the spinning drop and capsule compression experiments, 
the image analysis for radius and volume measurements, 
the SEM measurements of capsule thickness, the fit of the 
magnetic field distribution, and on the dynamic light scattering 
to determine magnetic nanoparticle size distributions.
List of chemicals.



\subsection{Shape equations}
\label{sec:shape_eqns}

In order to calculate axisymmetric shapes of capsules under the influence 
of external magnetic forces we numerically solve a closed set 
of six shape equations, which are based on nonlinear Hookean 
elasticity of the material. We recapitulate the shape equations in this
section briefly. 
  For more  details on the elastic model and the derivation 
of the shape equations, see Refs.\  \citenum{Knoche11,Knoche14} and 
Ref.\ \citenum{Wischnewski2018} in the presence of magnetic forces.
  
\begin{figure}
   \includegraphics[width=0.9\linewidth]{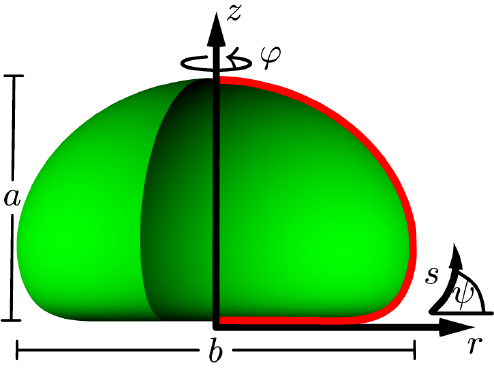}
 \caption{ 
  Parametrization of the axisymmetric capsule surface 
 in cylindrical coordinates. 
The red contour line is calculated numerically.  }
      \label{fig:capsule1_phases_supp}
\end{figure}

The capsules' shells are thin enough to be effectively treated as 
two-dimensional.  We parametrize the surface in cylindrical coordinates
$(r,z,\varphi)$ (see Figure \ref{fig:capsule1_phases_supp}). 
 Because of rotational symmetry, the contour line of the capsule
can be written as $z(r)$.  The arc length $s$ of the contour line starts at
the lower apex with $s = 0$ und ends at the upper apex with $s=L$.

We use a Hookean elastic energy density
\begin{align}
    \begin{split}
    w_S = \frac{1}{2}\frac{Y_{2\text{D}}}{1-\nu^2}
    (e_s^2+2\nu e_se_\varphi+e_\varphi^2)\\
	+\frac{1}{2}E_\text{B}(K_s^2+2\nu K_sK_\varphi+K_\varphi^2).
    \end{split}
\end{align}
The strains $e_i$ are related to the stretch factors $\lambda_i$ via $e_i =
\lambda_i -1$ and the bending strains $K_i$ are related to the curvatures
$\kappa_i$ via $K_i = \lambda_i \kappa_i - \kappa_{i_0}$.  The index $s$
describes the meridional direction and $\varphi$ the circumferential
direction. The index $0$ indicates the reference sphere.  The bending modulus
$E_\text{B}$ is defined via
  \begin{align}
    E_\text{B} = \frac{Y_{2\text{D}}h^3}{12(1-\nu^2)}.
  \end{align}
The constitutive relations 
  \begin{align}
    \tau_s &=\frac{Y_{2\text{D}}}{1-\nu^2}\frac{1}{\lambda_\varphi}
    (e_s+\nu e_\varphi),\\
    m_s &= E_\text{B}\frac{1}{\lambda_\varphi}(K_s+\nu K_\varphi)
  \end{align}
   for the ealstic stresses $\tau_i$ and the bending moments $m_i$
   ($\tau_\varphi$ and $m_\varphi$ with interchanged indices) are obtained by 
variation of the energy density with respect to the strains. 

The shape equations follow from purely geometric relations and 
the force and moment equilibrium in the shell, which is described 
by the following three equations (normal and tangential force equilibrium 
and moment equilibrium):
\begin{align}
    \label{eq:quilibrium}
      \begin{split}
      0 &= (\tau_s + \gamma)\kappa_s + (\tau_\varphi + \gamma)\kappa_\varphi \\
	  &\quad- (p_0 \Delta\rho gz + f_m)
    +\frac{1}{r}\frac{\text{d}(rq)}{\text{d}s},
      \end{split}\\
      0 &= \frac{\cos\psi}{r}\tau_\varphi+\kappa_sq
     -\frac{1}{r}\frac{\text{d}(r\tau_s)}{\text{d}s},\\
      0 &= q+\frac{1}{r}\frac{\text{d}(rm_s)}{\text{d}s}
       -\frac{\cos\psi}{r}m_\varphi.
  \end{align}
 Rearranging these equilibrium equations and geometrical relations 
   for $r$, $z$ and $\psi$, we get a system of six differential
 equations, the shape equations:
\begin{align}
    \label{eq:shape_eqns}
    r'(s_0)&=\lambda_s\cos\psi,\\
    z'(s_0)&=\lambda_s\sin\psi, \\
    \psi'(s_0)&=\lambda_s\kappa_s,\\
    \tau'_s(s_0)&=\lambda_s\left(\frac{\tau_\varphi-\tau_s}{r}\cos\psi
           +\kappa_sq-p_s\right),\\
    m'_s(s_0)&=\lambda_s\left(\frac{m_\varphi-m_s}{r}\cos\psi-q\right),\\
    \begin{split}
    q'(s_0)&=\lambda_s\biggr{(}-\kappa_s(\tau_s+\gamma)-\kappa_\varphi(\tau_\varphi+\gamma)\\
	    & \quad-\frac{q}{r}\cos\psi+\, p_0 + \Delta\rho gz + f_m \biggr{).} 
    \end{split}
   \end{align}
The first three equations are geometrical relations, while the remaining three
equations represent force and moment equilibrium.  
The pressure $p_0$ inside the capsule
is modified by the hydrostatic pressure $\Delta\rho gz$ and the magnetic
pressure $f_m$.  This system is closed by the following relations:
\begin{align}
    \lambda_s &= (1-\nu^2)\lambda_\varphi\frac{\tau_s}{Y_{2\text{D}}}-\nu(\lambda_\varphi-1)+1,\\
    \lambda_\varphi &= \frac{r}{r_0},\\
    K_s &= \frac{1}{E_\text{B}}\lambda_\varphi m_s-\nu K_\varphi,\\
    K_\varphi &= \frac{\sin\psi-\sin\psi_0}{r_0},\\
    \kappa_s &= \frac{K_s+\kappa_{s_0}}{\lambda_s},\\
    \kappa_\varphi &= \frac{\sin\psi}{r},\\
    \label{tauphi}
    \tau_\varphi &= \frac{Y_{2\text{D}}}{1-\nu^2}\frac{1}{\lambda_s}((\lambda_\varphi-1)+\nu(\lambda_s-1)),\\
    m_{\varphi} &= \frac{E_{\text{B}}}{\lambda_s}(K_\varphi+\nu K_s)
\end{align}
We solve the closed system of six 
 shape equations numerically with a fourth order Runge-Kutta
scheme. Boundary conditions at the two poles follow from the requirement 
of a closed capsule  and lead to a boundary value 
problem that is solved by employing  a multiple shooting method 
in conjunction with the Runge-Kutta scheme.

\subsection{Analysis of elastic parameters}

\begin{figure}[htb]
  \centering
  \includegraphics[width=1.0\linewidth]{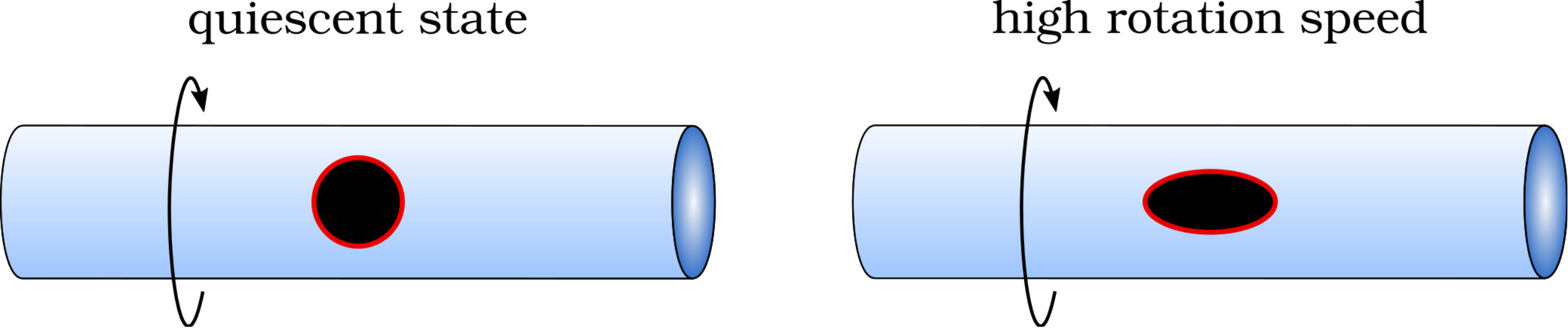}
  \caption{ Spinning capsule experiment.}
  \label{fig:SD}
  \end{figure}

For the spinning capsule experiments, we used the SVT~20 of the DataPhysics
Instruments GmbH. We used Fluorinert 70 (FC~70) as outer phase because of its
high density. The initial undeformed (quiescent) state was recorded at
2000~rpm.  In Figure \ref{fig:SD}, a sketch of the  spinning 
capsule measurement technique is shown.

\begin{figure}
  \centering
  \includegraphics[width=1.0\linewidth]{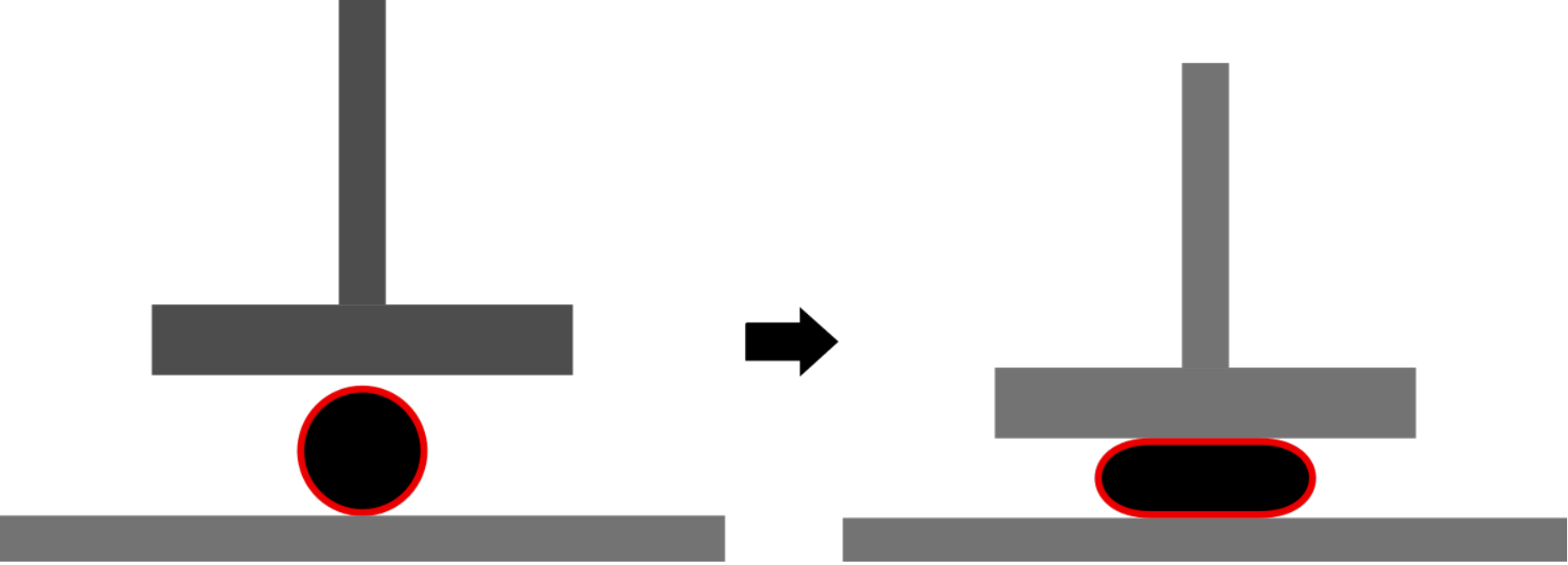}
  \caption{Capsule compression experiment.}
  \label{fig:CC}
  \end{figure}

Capsule compression experiments were performed with the DCAT11 tensiometer
(DataPhysics Instruments GmbH) with the respective software SCAT. The
compression speed was set to $0.02\,\mathrm{mm}/\mathrm{s}$. A sketch of the
capsule compression method is shown in Figure \ref{fig:CC}.

\subsection{Measurement of initial radius $R_0$ and volume $V_0$ of capsules}

The radius $R_0$ of the initial undeformed spheres can not be measured
directly, because the capsules are deformed by gravity even without an
external magnetic field.  Therefore, we determine
the capsules' volume from the experimental images.
 The image analysis was performed with FIJI/ImageJ
\cite{schindelin_fiji:_2012,rueden_imagej2:_2017}) 
using  capsule photos taken
with a OCA20 pendant drop tensiometer (DataPhysics Instruments GmbH).

In order to determine the volume $V_0$,  we measure the axial radius
 $r_i$ at different heights $z_i$ and calculate 
$V_0= \int dz \pi r^2(z)\approx \sum_i \pi r_i^2 (z_{i+1}-z_i)$ 
by summation over  small cylindrical volumes of radius $r_i$ at 
height $z_i$.
We then   calculate the initial radius
by $R_0 = (3/(4\pi)V_0)^{(1/3)}$.  We only assume
 that  capsules are axisymmetric, which is fulfilled to a good approximation.

We find that
the volume inside the capsule is constant during the whole experiment. The
elastic shell is impermeable for the involved fluids.

\subsection{Scanning electron
microscopy (SEM) images}

\begin{figure}
  \centering
  \includegraphics[width=1.0\linewidth]{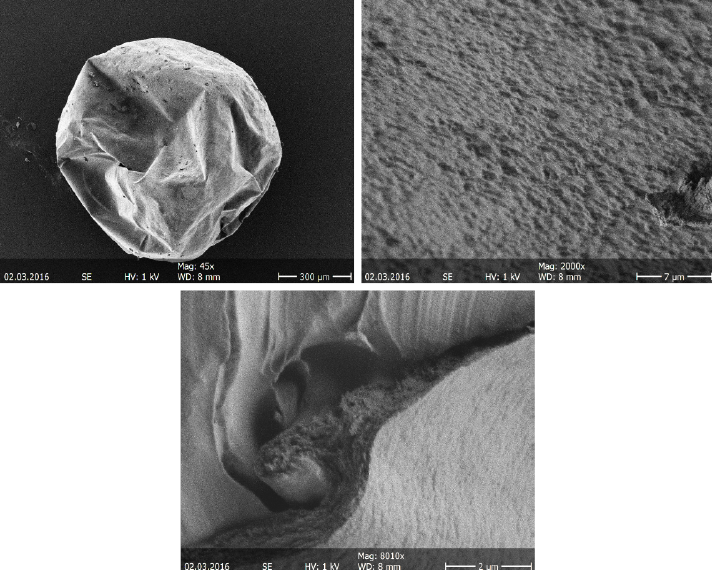}
  \caption{ 
  Scanning electron microscopy images of the ferrofluid-filled
    capsules. Center: complete capsule, bottom right: outer capsule shell.
    Top left, top right and bottom left: 
    cross sections through ruptured parts of the shell at increasing 
   megnification; yellow lines measure 
   shell thickness.}
  \label{fig:REM}
  \end{figure}

 In order to estimate the shell thickness, SEM measurements were
 performed. The resulting images are shown in Figure \ref{fig:REM}. The
 capsule was broken prior to the measurement to avoid bursting in ultra high
 vacuum.

  This gives  only a rough
  estimate  of the thickness of the capsule
  shell. In addition to potential errors due to optical effects like parallax
  error it has to be taken into account that SEM is performed in vacuum, i.e.,
   in the dried unhydrated state. With
  our optical microscope we were not able to resolve the hydrated 
   shell of  intact capsules, from  which we can
  conclude that the shell thickness of the hydrated capsules has to be
    below  $5\,\mathrm{\mu m}$.
  From the SEM images we find  a shell thickness of approximately
  $600\,\mathrm{nm}$ in vacuum in the dried state, see yellow lines in 
  Figure  \ref{fig:REM}.
  For alginate capsules with thicker shells we could perform 
  both optical microscopy measurements in the hydrated state 
   and SEM measurements in the dried 
   state. These measurements suggest swelling factors
   larger than  $5$ for the thickness increase by hydration. 
  We conclude that the shell thickness in the
  hydrated state is approximately $3\,\mathrm{\mu m}$ 
  with a relatively high error around $1\,\mathrm{\mu m}$.

\subsection{Fit of the external magnetic field}

The external magnetic field generated by a coil with a 
conical iron core was measured
with a hall probe.  The field was measured on different positions on the
central axis over the iron core and in the vicinity of  the axis.

\begin{figure*}
    \centering
    \includegraphics[width=.48\textwidth]{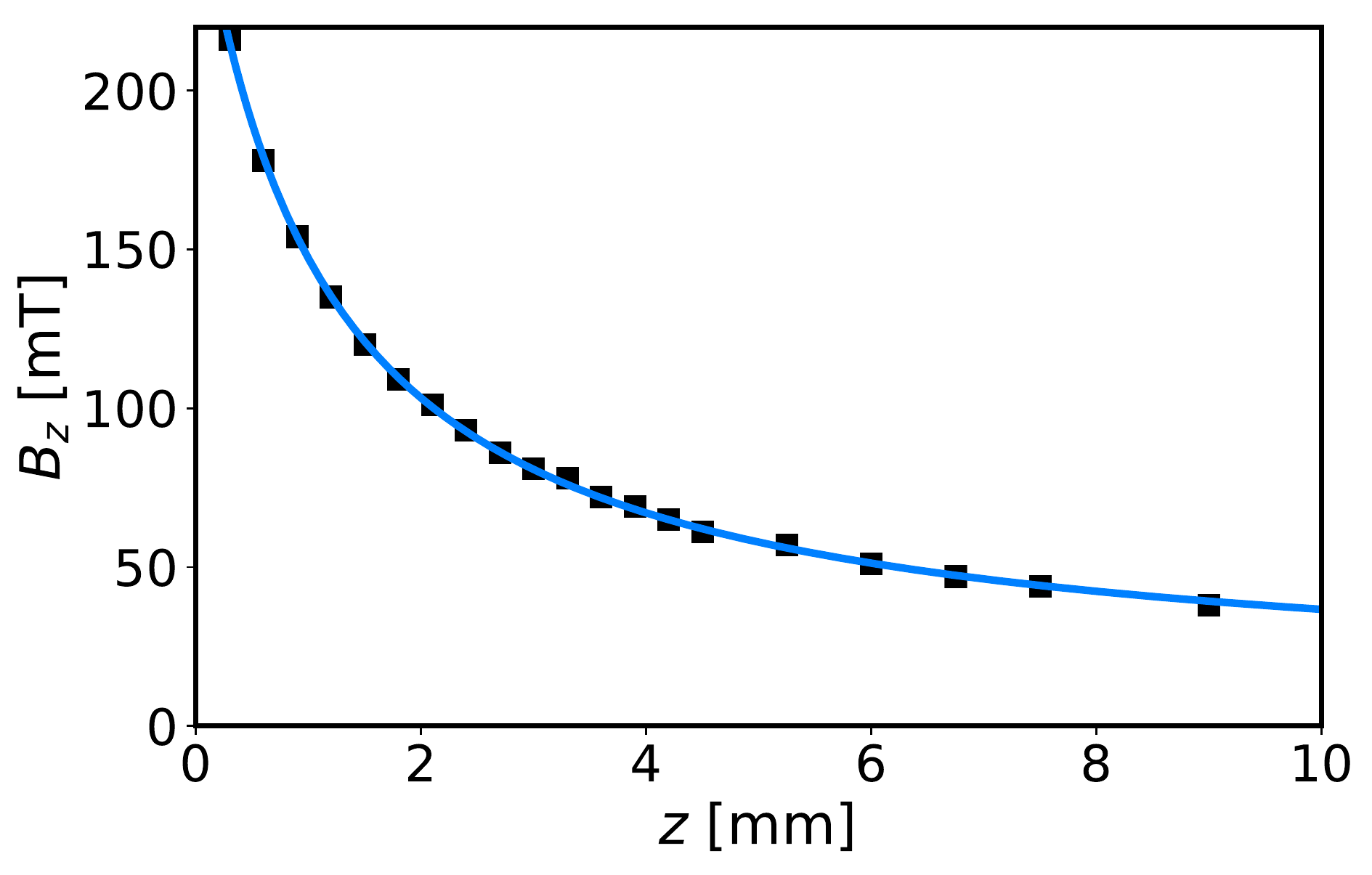}~
   \includegraphics[width=.48\textwidth]{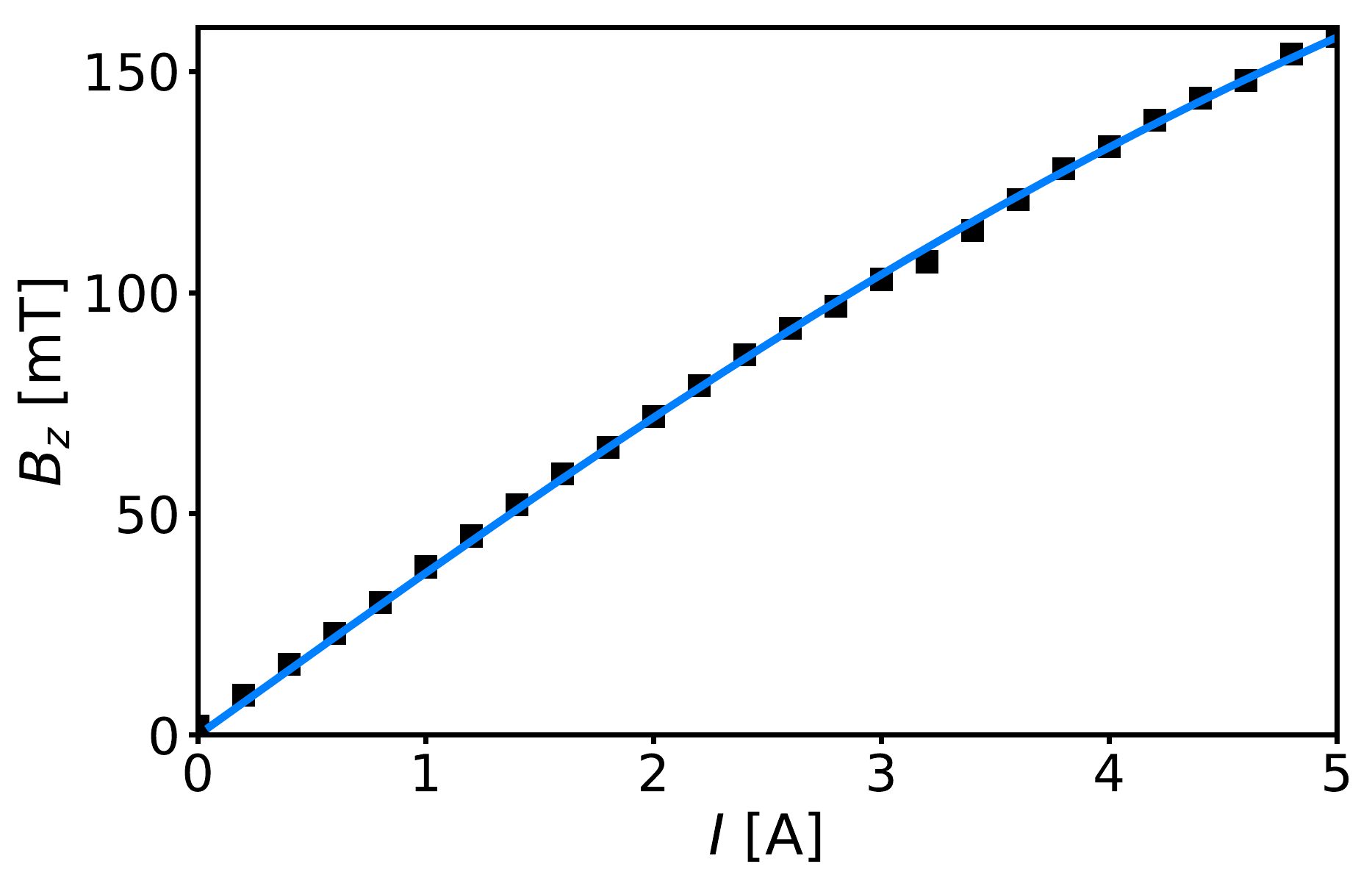}
    \caption{ 
     Magnetic flux density $B_z$ in $z$-direction as a function 
  of  the vertical distance $z$ 
      to the iron core for $I=2$\,A (left) and 
  as a function of the  current $I$ in the coil for  $z=3.6$\,mm 
  (right). 
   The fit eq \ref{eq:fieldfit} is shown as solid lines. 
 }
    \label{fig:field}
\end{figure*}

We found  that the 
field was nearly constant in radial direction within a distance of about
$2$\,mm next to the central axis.  Our biggest capsule had a radius of
$R_0=1.044$\,mm and even in the deformed state, its radial dimension 
 did not exceed $1.4$\,mm. 
 So it is well justified to treat the magnetic field as constant in
radial ($r$-) direction and to set the radial component of the field $B_r$ to
zero.  Together with the cylindrical symmetry of the setup, we only have to
estimate the $z$-component $B_z$ that depends on the coordinate $z$
and the current $I$.  We use a Langevin function to model the current
dependency and a hylerbolic function for the position dependency.  
The measured magnetic field is in good agreement with a fit
\begin{align}
      \label{eq:fieldfit}
      B_z(z, I) = a\left(\coth(b_I I) 
        - \frac{1}{b_I I}\right)\left(\frac{a_z}{z- b_z} + c_z\right)
\end{align}
with parameters 
    \begin{align}
      a &= 4.647 \\
      b_I &= 0.332\,\frac{1}{\text{A}}\\
      a_z &= 286.7\cdot 10^{-6}\,\text{Tm}\\
      b_z &= -1.104\cdot10^{-3}\,\text{m}\\
      c_z &= 10.86\cdot10^{-3}\,\text{T}.
\end{align}
  The fit describes the magnetic field with an error below 
1 \% in the neighborhood of the capsule.
  A plot of the $z$-  and  $I$-dependence 
 of the $B_z$  is shown in Figure \ref{fig:field}.

\subsection{Dynamic light scattering}

To ensure the stability of the nanoparticles inside the ferrofluid dynamic
light scattering measurements were performed. We used the ZetaSizer Nano ZS by
Malvern instruments with the Zetasizer Software 7.12. The results were
analysed with the CONTIN fit. All particle size values are given as the
so-called \emph{number mean}.

The capsules were produced by formation in a layer of distilled water followed
by sinking into layer of aqueous sodium alginate solution (1\%$_w$). The
polymerisation time needed for the formation of stable, ferrofluid-filled
capsules was 30~s. The procedure is shown in 
Figure 1 in the main text.

\subsection{Additional interface tension in capsule compression experiments}

The original Reissner formula
\begin{equation}
  F = \frac{4 Y_{2\text{D}} h}{R_0 \sqrt{3 (1-\nu^2)}} \: d
  \label{eq:reissner_supp}
\end{equation}
for the force-displacement relation 
describes the linear response  of an unpressurized and  initially 
  tension-free  elastic shell with rest radius $R_0$ 
  to a point force $F$ in terms of the resulting indentation $d$.
In the presence of 
 an additional interfacial tension, 
the main difference to the purely elastic shell is 
a non-vanishing pressure $p_0$, which is caused by 
the interfacial tension already in the initial state with $F=0$ 
and  which satisfies the Laplace-Young equation
\begin{align}
2\gamma/R_0 = p_0.
\label{eq:LY}
\end{align} 
We generalize Reissner's linearized shallow shell theory
\cite{reissner_stresses_1946,reissner_stresses_1946-1} 
to include the interfacial tension and the resulting internal pressure $p_0$ 
(in Ref.\ \citenum{Vella2012} the related problem of pressurized shells in the 
absence of an interfacial tension has been considered).
This leads to a linearized shallow shell equation
\begin{align}
\label{eq:Vella_32}
\kappa_B\nabla^4w-\gamma\nabla^2w+\frac{Y_{2\text{D}}}{R_0^2}w 
     = -\frac{F}{2\pi}\frac{\delta(r)}{r}
\end{align}
for the normal displacement $w(r)$ in polar coordinates,
with $r$ as the radial distance from the origin where the 
point force $F$ is applied; $\kappa_B$ is the shell's bending modulus  
\begin{align}
\kappa_B = \frac{Y_{2\mathrm{D}}h^2}{12(1-\nu^2)}.
\end{align}

Equation \eqref{eq:Vella_32} is identical to the linearized shallow shell 
equation governing the indentation of pressurized shells with internal 
pressure $p_0$ in the absence of interfacial tension 
(eq (3.2) in Ref.\ \citenum{Vella2012})
with the interfacial tension $\gamma=p_0R_0/2$ (according 
to Laplace-Young equation \ref{eq:LY})  replacing the 
pressure-induced isotropic stress $\sigma_\infty= p_0R_0/2$. 
This means both problems are equivalent: 
the interfacial tension gives rise to an internal  pressure $p_0$ 
in the same way as an internal pressure $p_0$ gives rise to 
an isotropic tension $\sigma_\infty$ prior to indentation. 
Consequently the solution of
eq \eqref{eq:Vella_32}  proceeds as in Ref.\ \citenum{Vella2012}.
Integrating eq \eqref{eq:Vella_32} 
The solution $w(r)$ for a given indentation $d = -w(0)$
has to be integrated over the whole reference plane of shallow 
shell theory to obtain the force $F = -2\pi (Y_{2\text{D}}/R_0^2)
   \int_0^\infty dr r w(r)$.
This  finally gives
\begin{align}
  F = \frac{4 Y_{2\text{D}} h}{R_0 \sqrt{3 (1-\nu^2)}} \: G(\tau)\:d
  \label{eq:reissner_modified_supp}
\end{align}
with
\begin{align}
  \label{eq:G(tau)}
  G(\tau) &= \frac{\pi}{2}
  \frac{(\tau^2-1)^{1/2}}{\operatorname{artanh}{(1-\tau^{-2})^{1/2}}}
 \\
  \tau &= 3(1-\nu^2)\left(\gamma/Y_{2\mathrm{D}} \right)^2\left(R_0/h \right)^2.
\nonumber
\end{align}
For $\gamma\approx 0$ the Reissner result  \eqref{eq:reissner_supp} 
is recovered
(note that in this case $\tau\approx 0$ and both the numerator and the 
 denominator in \eqref{eq:G(tau)} become imaginary because $0<\tau<1$).
For finite $\gamma>0$ we find a stiffening of the shell, i.e., 
an increased linear stiffness $F/d$ as in Ref.\ \citenum{Vella2012}
for pressurized shells ($G(\tau)>1$ for $\tau>0$). 
The increase in linear stiffness remains small for $\tau \ll 1$ 
($G(\tau) \approx 1$ for $\tau \ll 1$), which is fulfilled  
for capsule materials with $\nu$ sufficiently close to unity 
as for our alginate capsules. Therefore, corrections
due to $\gamma>0$ remain small.

\subsection{Additional interface tension in spinning drop experiments}

We also have to generalize the analysis of capsule deformation 
in the spinning drop apparatus given in Ref.\  
\citenum{pieper_deformation_1998} in the presence of 
interfacial tension. 
The linear response of the 
capsule deformation in spinning drop experiments 
is actually equivalent to the 
deformation of a ferrofluid-filled capsule in a  
small uniform external magnetic field as it has been 
analyzed in Ref.\ \citenum{Wischnewski2018} 
if we set the magnetic susceptibility to  
 $\chi = -1$ and the magnetic field strength to 
 $\mu_0H^2 = \Delta\rho R_0^2\omega^2$.
Both magnetic forces on the ferrofluid-filled capsule in an external 
field and 
  the centrifugal pressure exerted by a liquid of different density 
inside the capsule are normal forces (as they are generated by liquids)
acting on the capsule surface.
For a spherical shape the magnetic normal forces have the same 
position-dependence on the polar angle (between the radial 
vector and  the symmetry axis, i.e., field or 
rotation axis) for a susceptibility $\chi=-1$; 
for $\mu_0H^2 = \Delta\rho R_0^2\omega^2$
also the magnitude of magnetic and centrifugal forces becomes 
identical.  
In Ref.\ \citenum{Wischnewski2018}  the deformation of a 
ferrofluid-filled magnetic capsule has already been considered 
also in the presence of an interfacial tension $\gamma$, and we 
can simply adapt the results for the linear response, which are 
derived in Appendix A of  Ref.\ \citenum{Wischnewski2018},
to the  capsule deformation 
in the spinning drop apparatus by exploiting the equivalence 
of both problems.

For completeness we repeat  the major steps of the calculation.
The linear response theory is of first order in the 
displacements ($u_r, u_\varphi, u_\theta$) in spherical coordinates 
with the polar angle $\theta$ and the azimuthal angle $\varphi$ and 
$\theta=0$ as the rotation axis 
(note that in Ref.\ \citenum{pieper_deformation_1998} the notation is 
different: the azimuthal angle 
is denoted by $\theta$ and the polar angle by $\phi$).
In the elastic shell, the equilibrium of forces has to be 
fulfilled in tangential and normal direction.
The tangential force equilibrium is given as
\begin{align*}
\frac{\mathrm{d}}{\mathrm{d} \theta}(R_0\tau_{\theta} \sin \theta) 
  =  R_0 \tau_\varphi\cos\theta
\end{align*}
and the normal equilibrium as
\begin{align*}
\begin{split}
\frac{1}{R_0}(\tau_\theta + \tau_\varphi) 
   + (\kappa_\theta + \kappa_\varphi)\gamma\\
 = p_0 + \frac{1}{2}\Delta\rho R_0^2\omega^2\sin^2\theta.
\end{split}
\end{align*}
The curvatures $\kappa_\varphi$ and $\kappa_\theta$ are expanded 
to first order in the displacements via
\begin{align*}
\kappa_\theta + \kappa_\varphi \approx \frac{2}{R_0} 
- \frac{1}{R_0^2}(2u_r - \partial_\theta^2u_r + \partial_\theta u_r\cot\theta).
\end{align*}
The coupled equation system describing force balance 
has to be solved using the constitutive relations
\begin{align*}
\tau_\varphi - \nu\tau\theta &= 
   \frac{Y_{2\mathrm{D}}}{R_0}(u_\theta\cot\theta + u_r)\\
\tau_\theta - \nu\tau\varphi &= 
  \frac{Y_{2\mathrm{D}}}{R_0}(\partial_\theta + u_r)
\end{align*}
and the boundary conditions 
$\partial_\theta u_r(0)=\partial_\theta u_r(\pi/2)=0$ 
and $u_\theta(0)=u_\theta(\pi/2)=0$.
The ansatz
\begin{align*}
u_r &= A + B\cos^2\theta\\
u_\theta &= C\sin\theta\cos\theta,
\end{align*}
 describes a spheroidal shape, 
preservation of volume requires $A = -B/3$.
Using this spheroidal ansatz we find the solution
\begin{align*}
A & = \frac{\Delta\rho R_0^4\omega^2(5+\nu)}
      {24[Y_{2\mathrm{D}} + (5+\nu)\gamma]}\\
B & = -3A\\ 
C & = \frac{\Delta\rho R_0^4\omega^2(1+\nu)}
{4[Y_{2\mathrm{D}} + (5+\nu)\gamma]}.
\end{align*}
To calculate the deformation parameter $D$, we use 
$r(\theta) = R_0 + u_r$ and
find 
\begin{align}
D &= \frac{r(0) - r(\pi/2)}{r(0) + r(\pi/2)} 
\nonumber\\
  &\approx \frac{B}{2R_0} 
= \frac{-\Delta\rho R_0^4\omega^2(5+\nu)}{16[Y_{2\mathrm{D}} +
    (5+\nu)\gamma]}.
\label{eq:SD_supp}
\end{align}
For $\gamma \approx 0$ we recover the 
well-known result from Ref.\ \citenum{pieper_deformation_1998},
\begin{equation}
  D = -\Delta \rho \omega^2 R_0^3 \frac{(5 + \nu)}{16 Y_{2\text{D}}},
  \label{eq:SD_original_supp}
\end{equation}
This means that, in  the presence of an interfacial tension $\gamma>0$, 
 we simply have to 
replace $Y_{2\text{D}}$ in eq \ref{eq:SD_original_supp} by 
$Y_{2\text{D}}  + (5+\nu)\gamma$ resulting in a reduction of the 
deformation  $D$.
Analyzing the same experimental spinning capsule 
deformation data should give identical values of $Y_{2\mathrm{D}} +
    (5+\nu)\gamma$. Analyzing the same data assuming  $\gamma>0$ will 
thus reduce the inferred result for the Young's modulus
by $(5+\nu)\gamma$. 

Including an interfacial tension $\gamma$ into the combined analysis of 
spinning drop and capsule compression experimental data 
thus results in a reduced elastic modulus $Y_{2\mathrm{D}}\to 
 Y_{2\text{D}}  - (5+\nu)\gamma$  to fit the spinning drop 
data with eq \eqref{eq:SD_supp}. 
If $\gamma/Y_{2\text{D}} \ll 1$ such that $\tau\ll 1 $ in 
 eq \eqref{eq:reissner_modified_supp}, 
 $\nu$ is increased  at the same time such that 
$Y_{2\mathrm{D}}/\sqrt{1-\nu^2}$ remains approximately unchanged to 
fit the capsule compression data with eq \eqref{eq:reissner_modified_supp}.

\subsection{Sensitivity of the capsule deformation to the Poisson number $\nu$}

\begin{figure}
  \centering
  \includegraphics[width=1.0\linewidth]{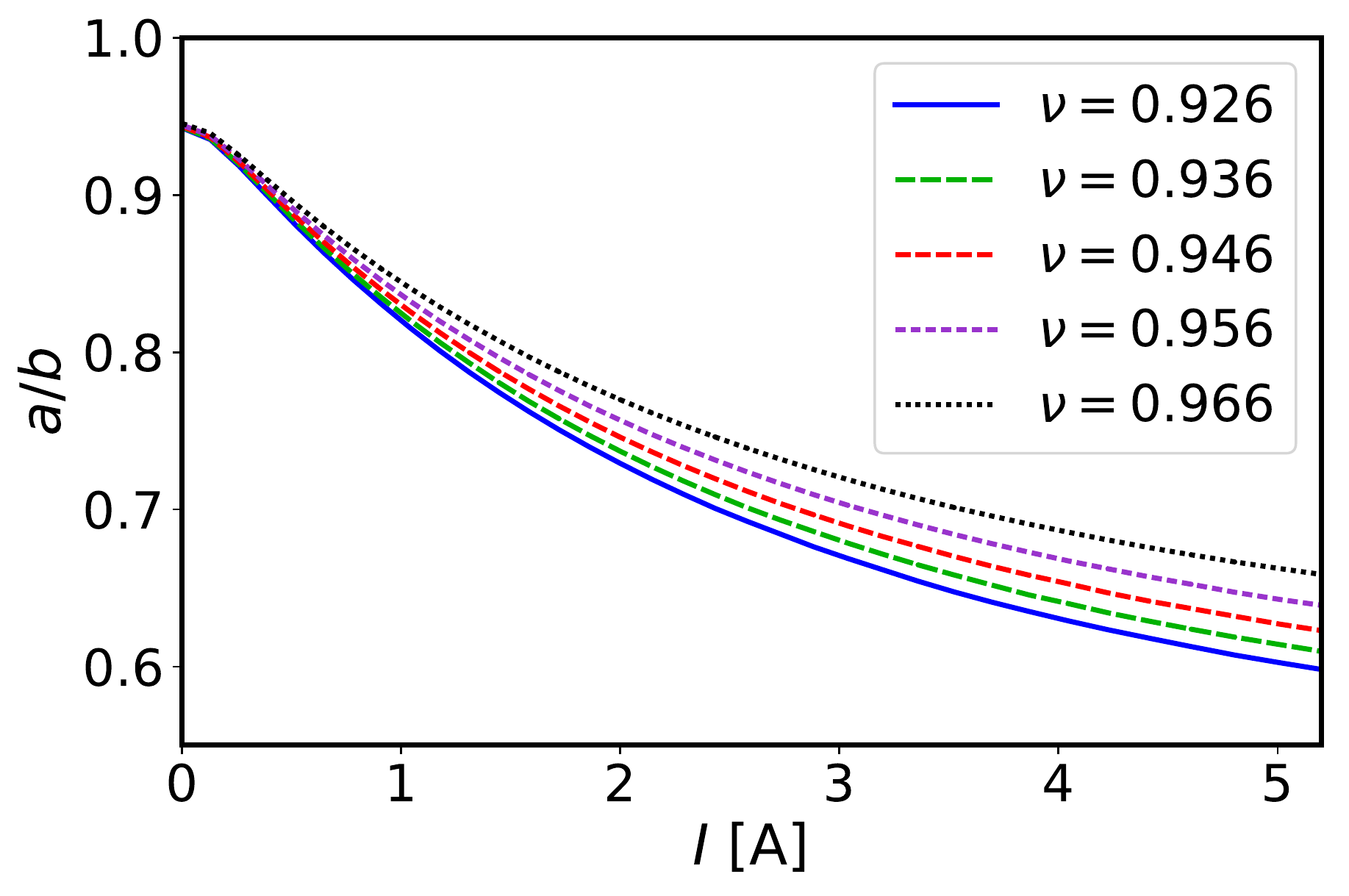}
  \caption{ 
      Ratio of the height $a$ to  width $b$ of capsule 1 for
      increasing current $I$ with $\nu = 0.926, 0.936, 0,946, 0.956, 0.966$.
    }
  \label{fig:Poisson_Sensitivity}
\end{figure}

A Poisson ratio close to unity gives rise to a large area compression 
modulus  $K_{2\text{D}} = Y_{2\text{D}}/2(1-\nu)$, which becomes sensitive to 
changes in $\nu$ and, thus, also large  elastic stresses, 
which contain  factors $1/(1-\nu)$ and 
 become  very sensitive to changes in $\nu$ (see prefactors 
 $1/(1 - \nu^2)$  in the elastic stresses in eq \ref{tauphi}.
Small deviations in $\nu$ lead to considerable changes in $\tau_i$ and 
visibly changing shapes of the capsule. 
Figure \ref{fig:Poisson_Sensitivity} shows simulations of 
capsule 1 with five 
slightly different Poisson ratios ranging from $\nu = 0.926$ to $\nu =0.966$. 
While there are only minor deviations in the weakly 
deformed state in the absence of magnetic fields ($I = 0\,$A), where 
the capsule is only  deformed by gravity and tensions are 
dominated by the $\nu$-independent interface tension, 
the capsule's side ratio $a/b$ shows obvious differences 
in the strongly deformed state ($I = 5\,$A), 
which is dominated by large elastic tensions.
In conclusion, strongly deformed shapes are 
better suited to determine the value of $\nu$. 
We can estimate the error of the 
numerically determined value of $\nu$ to be smaller than  $0.01$.

\subsection{List of chemicals}

  All chemicals including purity are shown in table \ref{tab:Chem}. All
  chemicals were used without further purification.
  \begin{table}[h]
  \caption{List of chemicals}
  \begin{footnotesize}
  \begin{center}
  \begin{tabular}{llc}
  Chemical & Manufacturer & Quality \\ \hline
  Calcium chloride anhyd. & Merck & 98\% \\ 
  Chloroform & Merck & p. A. \\ 
  Diphenyl ether & Acros Organics & 99\% \\ 
  Ethanol & VWR & 99,5\% \\ 
  Fluorinert (FC-70) & abcr & - \\ 
  $n$-Hexane & Merck & $\geq$96\% \\ 
  1-Hexanol & Merck & $\geq$ 98\% \\ 
  1,2-Hexadecanediol & Sigma Aldrich & 90\% \\ 
  Iron acetyl acetonate & Alfa Aesar & - \\ 
  Oleic acid & VWR & 81\% \\
  Oleylamine & Sigma Aldrich & 70\% \\ 
  Sodium alginata & Aldrich & - \\ 
  Sodium chloride & VWR & - \\ 
  \end{tabular}
  \end{center}
  \end{footnotesize}
  \label{tab:Chem}
  \end{table}

\bibliography{literature}

\end{document}